\documentclass[10pt,letterpaper]{article}
\usepackage[top=0.85in,left=1.75in,footskip=0.75in,marginparwidth=2in]{geometry}

\usepackage{array}
\usepackage{subfig}
\usepackage{amsmath,amssymb,amsfonts,amsthm}

\usepackage{url}
\usepackage{color}
\usepackage[utf8]{inputenc}
\usepackage[T1]{fontenc}
\usepackage{nicefrac}
\usepackage{subfig}
\usepackage{comment}
\usepackage{colortbl}

\usepackage{graphicx}

\usepackage{soul, color}

\newlength{\textlarg}

\def\bydef{\mathop{\overset{\mbox{\scriptsize def.}}{=}}\nolimits}
\def\one{{\mathchoice {\rm 1\mskip-4mu l} {\rm 1\mskip-4mu l}
{\rm 1\mskip-4.5mu l} {\rm 1\mskip-5mu l}}} 
\def\R{\mathbb{R}}
\def\X{\mathcal{X}}
\def\V{\mathcal{V}}
\begin{document}
\vspace*{0.35in}

\begin{flushleft}
{\Large
\textbf\newline{A new Edge Detector Based on Parametric Surface Model:\\Regression Surface Descriptor.}
}
\newline
\\ 
R\'emi Cogranne$^\star$, R\'emi Slysz, Laurence Moreau$^\dag$ and Houman Borouchaki$^\dag$\\
\bigskip
$\star$ Troyes University of Technology, Lab. for System Modelling and Dependability,\\
ROSAS Dept., ICD, UMR 6281 CNRS, Troyes 10010 Cedex, CS 42060, France
\\
$\star$ Lab. for System Modelling and Dependability
$\dag$ UTT - INRIA joint research unit GAMMA3 - Automatic mesh generation and advanced methods
\\
\bigskip

\end{flushleft}

\section*{Abstract}




In this paper we present a new methodology for edge detection in digital images. The first originality of the proposed method is to consider image content as a parametric surface. Then, an original parametric local model of this surface representing image content is proposed. The few parameters involved in the proposed model are shown to be very sensitive to discontinuities in surface which correspond to edges in image content. This naturally leads to the design of an efficient edge detector. Moreover, a thorough analysis of the proposed model also allows us to explain how these parameters can be used to obtain edge descriptors such as orientations and curvatures.\\
In practice, the proposed methodology offers two main advantages. First, it has high customization possibilities in order to be adjusted to a wide range of different problems, from coarse to fine scale edge detection. Second, it is very robust to blurring process and additive noise. Numerical results are presented to emphasis these properties and to confirm efficiency of the proposed method through a comparative study with other edge detectors.

\section*{Keywords}
Image edge detection ;
Content analysis ; 
Feature extraction ; 
Image representation ; 
Parametric model.

\section{Introduction}\label{sec:intro}

Edge detection consists in locating significant local changes of image intensity usually associated with a discontinuity in the image content. It is a fundamental problem of image processing and computer vision which has been widely studied. In fact, identifying and locating edges is a low-level task in a wide range of applications
such as compression, features extraction, pattern recognition, tomography defect detection, image restoration and enhancement,  etc.
For many years, a vast majority of computer vision methods have used edge detection as a pre-process to simplify natural image descriptor while preserving the structured content.
The reader may find a detailed introduction on edge detection in~\cite[Chap.5]{Jain95}.

\subsection{State-of-the-Art}

In the literature, many different edge detection methods have been proposed. They can roughly be divided into the following main categories:
\begin{itemize}
  \item Algorithms for edge detection relying on a linear high-pass filtering of the analyzed image, usually in order to compute discret gradient, are probably the most popular~\cite{Canny86,Haralick84,Huertas86,Prewitt,Sobel}. These linear high-pass filters are numerically implemented as a discrete convolution, usually using a kernel of small size. Therefore filtering approach provides overall good quality edge detectors with computationally low complexity.
  \item Multi-scale analysis, with its successful introduction in image processing through wavelet decomposition, has also been widely used to design edge detector algorithms~\cite{Lindeberg96,Sheng09,Zhang02}. Due to the nature of multi-scale analysis, these methods fall in the filtering approach, as the underlying methodology essentially consists in an iterative application of a discrete convolution. 
  \item The variational approach has also lead to the design of efficient edge detectors~\cite{MorelLionsAll92,Perona90}. Roughly speaking, the underlying theory is that an image can be represented as an object which evolves in time, from observed scene to acquired image, by application of successive convolution operations. Though the variational approach leads to the design of very efficient edge detectors, its major drawback is a high computational cost due to iterative research of an inverse problem solution.
  \item Nonlinear image processing methods have been extensively used for purposes closely related to edge detection, like edge-adaptive image denoising, interpolation, deblurring, etc. The reader may refer to~\cite{Hardie95,Mitra2001Nonlinear,Raez84} and the references therein. In the field of edge detection, the nonlinear approach have been mainly used through median filtering~\cite{Bovik87} but other methods have been proposed, see~\cite{Laligant2010,Pitas86} for instance.
\end{itemize}

It should be denoted that an exhaustive review of edge detection methods is hardly possible as in the literature, this problem have been widely studied during the three last decades. Almost each novel image processing method have been applied to address the edge detection problem. For instance, a method based on sparse dictionary learning is proposed in~\cite{Mairal08}. Finally, an interesting paper \cite{Unser1993} should be mentionned as it has a similar approach to the presented method in the way that it is based on magnitude of B-spline approximation coefficients.

\subsection{Contribution and Organization of this Paper}
The main originality of the proposed edge detection method lies on the fact that the image is locally considered as a parametric surface. A simple parametric model using a multidimensional polynomial is presented to locally approximate this surface. The few parameters involved in the proposed local model are shown to be very sensitive to the presence of an edge or a surface high gradient. This naturally permits the design of a simple yet efficient edge detection algorithm. It is also shown that an analysis of the parameter used to locally approximate surface allows the extraction of edge descriptors and the detection of particular edges (with a desired orientation or curvature). Finally, the proposed methodology depends on the polynomial order used to locally approximate image content as a surface and on the  window size on which image is analyzed. 
These two parameters provide a high level of customization of the proposed method; for instance, a coarse or a fine edge detection is possible as well as robustness with respect to blur or additive noise.

The contribution of the edge detection method proposed in this paper can be sum up as follows. First, we modeled image content as a locally parametric surface described by few parameters which are shown to be sensitive to the presence of edges. Second, the presented methodology can be used in a general framework for edge detection but also to extract local edge descriptors and retrieve specific edge patterns. Third, the proposed methodology is shown to be robust to noise addition and blurring process. 

This paper is organized as follows. Section~\ref{sec:image_model} presents a generic model of a natural image content based on light emission and imaging device properties. Based on this image content model, the proposed method for edge detection, called RSD for \emph{Regression Surface Descriptor}, is detailed in Section~\ref{sec:proposed_method}. It is particularly shown that the proposed method can be extended to extract edge descriptors, such as orientation and curvature, and to detect only edges with desired properties. 
Finally, Section~\ref{sec:num_results} emphasis the relevance of the proposed method through numerical results and comparison with other edge detectors. 

\section{Underlying Image Model and Edge Definition}\label{sec:image_model}

\subsection{From Imaged Scene to Recorded Image}\label{subsec:image_model_scene}

Let the imaged scene be defined as a function $S(x,y,\lambda)$ which represents the incident spectral radiance on photo-sensor. The variables $(x,y)$ represent the (continuous) coordinates of points defined on the bounded compact set $\X = [0,x_M]\times [0,y_M]$ hence $(x,y)\in \X \subset \R^2$. 
For each color $c$ (usually $c =\{r,g,b\}$), the ``effective channel radiance'' $S_c(x,y)$ is the spectral filtered version of $S(x,y,\lambda)$ given as~\cite{Healey94}:
\begin{equation}\label{eq:eff_irradiance}
  S_c(x,y) = \int_{\R^+} T_c(\lambda) S(x,y,\lambda) \, d\lambda
\end{equation}
where the spectral transmittance $T_c(\lambda)$ accounts for color filter, photon energy, and sensor quantum efficiency (number of electron per incident photon).
In the rest of this paper color channels are processed individually, therefore the index $c$ is omitted.
The function $S(x,y)$ is introduced for clarity and has a physical meaning only from a camera point of view.

\begin{figure*}[!ht]
\centering
\renewcommand{\arraystretch}{0}
\begin{tabular}{c c c}
\subfloat[Close-up on Lena's eye.]{
  \includegraphics[width=0.27\textwidth]{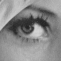}
  \label{fig:Lena_eyesOrig} }
  &
\subfloat[Close-up on Lena's eye displayed as a surface mesh.]{
  \includegraphics[width=0.34\textwidth]{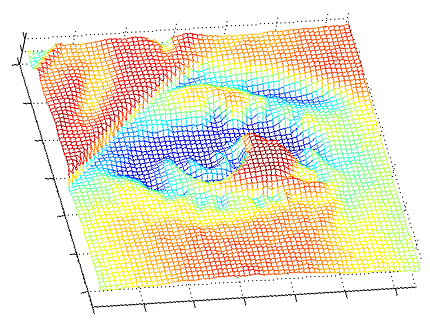}
  \label{fig:Lena_eyesMesh} }
  &
\subfloat[Close-up on Lena's eye approximation using proposed surface model for regression.]{
  \includegraphics[width=0.27\textwidth]{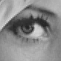}
  \label{fig:Lena_eyesApprox} }
\end{tabular}
\caption{Illustration of the proposed methodology: Lena's eye is shown on the left figure, representation as a discrete surface is shown on the middle figure and its approximation using the proposed efficient 2D polynomial model is illustrated on the right (shown as an image for display purpose).}
\label{fig:Lena_eyes}
\end{figure*}

The model of $S(x,y)$ used in this paper is inspired from~\cite{Cog2012TMI,MumfordShah89}. 
A scene is assumed to be made of $N_S$ solid objects~; hence the scene radiance $S(x,y)$ can be represented as piecewise continuous:
\begin{equation}
\label{eq:def_scene_2D}
 S(x,y) = \sum_{i=1}^{N_S} S_i(x,y)\,\one_{\X_i}(x,y)
\end{equation}
where $S_i(x,y)$ are continuous real functions over $\X$, $\one_{A}$ is the indicator function of the set $A\subset\R^2$, $\{\X_i\}$ is a finite partition of $\X$, each atoms $\X_i$ is connected with a nonempty interior $\mathring{\X}_i$ and a zero Lebesgue-measure boundary $\partial \X_i$. The discontinuity set of $S(x,y)$ is denoted $\Gamma \subset \bigcup_{i=1}^{N_S}\partial \X_i$. It is assumed that $\Gamma$ is a parametrized curve. This means that the discontinuity set $\Gamma$ can be considered as a piecewise continuous function $\Gamma:~T\mapsto \X$, where $T$ is a compact interval in $\R$, which admits the following representation:
\begin{equation}
\label{eq:curve}
 \Gamma(t) = \sum_{i=1}^{N_C} \gamma_i(t)\,\one_{T_i}(t)
\end{equation}
where the $\gamma_i's$ are the pieces of parametric curves that form $\Gamma$, $\{T_i\}$ is a finite partition of $T$ and $T_i$ is a nonempty segment. Less formally, this means that the set of discontinuities of $S(x,y)$ can be represented as the juxtaposition of a finite number of segments. To allow a simple mathematical formulation, it is supposed that the sets $\{\gamma_i(t)\,,\:t\in T_i\}$ are separated.
The image model described by~(\ref{eq:def_scene_2D}) and~(\ref{eq:curve}) is sufficiently general to describe a large class of natural scenes (see details in~\cite{MumfordShah89}).

The scene $S(x,y)$ is then acquired with an imaging device. Camera optical system can be modeled by a point spread function (PSF), denoted $h(\cdot)$ which mathematically represents a convolution kernel. Modeling the PSF of a camera is a difficult task because it depends on many elements from the acquisition pipeline~\cite{Goodman05}. Without loss of generality, the impact of camera PSF on recorded image is given by:
\begin{equation}
\label{eq:image_ideal_optic}
  I(x,y) = S \ast h {\bydef}\iint_{\mathbb{R}^2}\!\!S(x{-}u,y{-}v) h(u,v;x,y) \, du \, dv
\end{equation}
where $\ast$ represents the two-dimensional convolution product defined in~(\ref{eq:image_ideal_optic}).

From the definition, given in Equation~(\ref{eq:eff_irradiance}), of the ``effective channel radiance''~$S(x,y)$, $I(x,y)$ represents the irradiance incident on photo-sensor plane expressed as the number of photo-electrons generated per unit area at location $(x,y)$. It should also be noticed that the PSF is denoted $h(u,v;x,y)$ at location $(x,y)$ because it generally has a non-stationary property in the sense that this function depends on considered coordinates.

\subsection{Model of Recorded Image}\label{subsec:image_model_image}

As highlighted in~\cite{MumfordShah89}, one can expect from the physical properties of solid objects light emission that:
\begin{description}
 \item[P-1] Over each domain $\X_i$, associated with solid object $N_i$, emitted radiance varies smoothly.
 \item[P-2] The radiance is discontinuous across (most of) the boundaries $\gamma_{i}$.
 \item[P-3] Objects are of regular shape in the sense that each curves $\gamma_{i,j}$ are regulars, or $\gamma_{i}\in C^2(T_i)$.
\end{description}

In the present paper it is proposed to locally model image content. 
Under some mild conditions, it can be proven that the scene radiance $S$ can locally be represented, in a neighborhood $\V_{x_0,y_0}$ of $(x_0,y_0)\in\X$, as follows~\cite{Cog2011IH,Cog2012TIP}:
\begin{equation}\label{eq:model_scene_decomp}
  \forall (x,y) \in \V_{x_0,y_0} \,,\, S(x,y) = S_c(x,y) + S_d(x,y)
\end{equation}
where $S_c$ is a continuous function, $S_d(x,y)$ is a 2D step function which represents the discontinuity of the scene:
\begin{equation}\label{eq:model_scene_decomp_disc}
  S_d(x,y) = d(x,y) \one_{(x,y)\in\mathcal{X}_j}
\end{equation}
and $d(x,y)$ is the local intensity of radiance's discontinuity between domains $\X_i$ and $\X_j$.
 
>From the linearity of convolution product, the incident irradiance on photo-sensor plane~(\ref{eq:image_ideal_optic}) can be written:
\begin{equation}\label{eq:model_image_decomp}
  \forall (x,y) \in \V_{x_0,y_0} \,,\, I(x,y) = I_c(x,y) + I_d(x,y)
\end{equation}
where $I_c$ and $I_d$ respectively represent the convoluted continuous and discontinuous part, $I_c = S_c \ast h$ and $I_d = S_d \ast h$.

From properties~[P-1]-[P-3] and Equations~(\ref{eq:def_scene_2D})-(\ref{eq:model_scene_decomp}), the continuous part of incident irradiance $I_c(x,y)$ is expected to varies very slowly. Therefore, the gradient and Hessian of function $I_c$ are expected to have a small magnitude. On the contrary, the discontinuity part of irradiance is essentially a discontinuity blurred by imaging device's acquisition process. Hence, along the curves of discontinuity $\{\gamma_i\}$, the gradient and Hessian of function $I_d$ are expected to have a large magnitude. Note that the function $S_d$~(\ref{eq:model_scene_decomp_disc}) is constant outside the discontinuity set, therefore, function $I_d$ can be considered as constant outside blurred edges.

\begin{figure}[t]
\centering
\begin{tabular}{@{}c@{}c}
\subfloat[A very simple edge.]{
    \includegraphics[width=0.22\textwidth , trim=6 6 6 6 , clip]{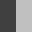}
  \label{fig:testimage} }
  &
\subfloat[Evolution map of the absolute value of regression coefficient $c_1$ for a 11x11 neighborhood. The hightest values are shown in black and white areas represents zero.]{
    {\setlength\fboxsep{0pt}
    \fbox{\includegraphics[width=0.22\textwidth , trim=6 6 6 6 , clip]{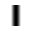}}}
  \label{fig:testimagec1} }
\end{tabular}
\caption{A trivial edge example.}
\label{fig:simple_test}
\end{figure}

More formally, the following trivial example provides an insight about the proposed model of scene and ensuing recorded image. Let assume that the imaged scene is made of two solid objects with constant radiance separated by the curve of Equation $x=x_0$, see Figure~\ref{fig:simple_test}. Moreover, let assume that the optical system of imaging device has a 2D Gaussian isotropic PSF:
\begin{equation}\label{eq:PSF}
h\left(u,v;s(x,y)\right) = \frac{1}{2\pi s^2(x,y)}\exp\left(-\frac{u^2+v^2}{2 s^2(x,y)}\right)
\end{equation}
where $s(x,y)>0$ is the local blur parameter at the considered image point $(x,y)$. Note that $s(x,y)$ has very smooth variations. Thus, we assume that the derivative is locally negligible.

By using properties~[P-1]-[P-3], a straightforward calculation allows the writting of the following expression of the imaged scene:
\begin{equation*}
I(x,y) = s_1 + (s_2 - s_1) \Phi\left( \frac{x-x_0}{s(x_0,y)} \right)
\end{equation*}
where $s_1$ and $s_2$ respectively represent the radiance over domains $\X_1$ and $\X_2$, and $\Phi(x) = (2\pi s^2(x,y))^{-1/2} \int_{-\inf}^x(\exp(-u^2/(2 s^2(x,y))) du$ represents the integral of Gaussian function with blur parameter ${s(x_0,y)}$. Hence, a Taylor expansion of function $I$ around a point $(x,y)$ immediately permits the writting of:
\setlength{\arraycolsep}{0.5em}
\begin{align}\label{eq:edge_Taylor}
  I(x\!+\!\eta,y) &\approx s_1 + (s_2 - s_1) \Phi\left( \frac{x-x_0}{s(x,y)} \right) \\
  &+ \eta\, (s_2 - s_1) \frac{1}{\sqrt{2\pi} s(x,y)} \exp\left( -\frac{1}{2}\frac{(x-x_0)^2}{s^2(x,y)} \right)    \nonumber    \\
  &- \eta^2\,  \frac{(x-x_0) (s_2 - s_1)}{2\sqrt{2\pi}s^3(x,y)} \exp\left( -\frac{1}{2}\frac{(x-x_0)^2}{s^2(x,y)} \right)               \nonumber
\end{align}

Equation~(\ref{eq:edge_Taylor}) clearly highlights that as $\lvert x-x_0 \rvert$ increases, coefficients of the Taylor serie expansion rapidly decrease at a rate $o(\exp(-\lvert x - x_0 \rvert)$. On the contrary, the first order coefficient reaches a maximum on $x=x_0$. Similarly, Equation~(\ref{eq:edge_Taylor}) shows that the coefficient of Taylor polynomial expansion linearly depends on $(s_2 - s_1)$, which represents the intensity of discontinuity. Finally, it can be noted that the polynomial coefficient decrease at the rate of $s(x,y)^{-p}$; this shows a direct consequence of Gaussian blur, the more important the blurring parameter is, the fewer coefficients are, because the edge tends to become smoother.

However, it should be highlighted that the blurring process also affects the continuous part of irradiance in the same way than the edges. 
Especially, it is assumed that the function $I_c$ can accurately be locally modeled using a Taylor expansion, see Figure~\ref{fig:Lena_eyes}. Hence, a direct consequence is that if an edge has higher Taylor expansion polynomial coefficients than the part $I_c$, this order relation holds after application of a stationary blurring process (with a PSF that is constant on image domain $\X$).

\section{Proposed Method: RSD}\label{sec:proposed_method}

The proposed method for edge detection is based on the irradiance model described in Sections~\ref{subsec:image_model_scene} an~\ref{subsec:image_model_image}. The idea is to seek for slightly blurred discontinuities in the radiance function ---representing pixels value--- of the acquired digital image by analysing coefficients of the Taylor series expansion. It should be noted that the  orientation and curvature of edges greatly influence those coefficients, see the example provided in Section~\ref{subsec:image_model_image}. The reader is referred to~\cite{Haralick84} in which those coefficients are used to estimated the most likely edge orientation and then to compute zero crossing second derivatives. To allow a simple and reliable detection of edges, the method proposed in this paper is directly based on the coefficients of the Taylor series expansion. In fact, it is likely that avoiding complicated intermediate steps improves the reliability of the edge detection method.

The method proposed in the present paper rely on the fact that whatever the geometrical properties of edges might be, the polynomial coefficients are likely to be non zero around edges. This assumption, detailed in Section~\ref{subsec:image_model_image}, allows us to propose a simple yet efficient edge detector based on the magnitude of those coefficients. In addition, a deeper analysis of those coefficients can be used for finding edges with specific orientation and curvatures, see Section~\ref{sec:edge_description}.

By using a Taylor expansion to locally model one channel of the image\footnote{Note that only the RGB encoding have been studied in this paper because it is possible to represent any image in this color space.}, it is possible to construct a method with no pre-conceived idea on the searched edges shape contrarily to high-pass filters category of edge detectors, see Section~\ref{sec:intro}. The image content model described in Section~\ref{sec:image_model} brings light on what kind of area could contain an edge. Translating in discrete surface representation, the properties [P-1]-[P-2] represent the well-known properties that areas with low gradient does not contains edges.

It has also been shown that the discontinuity part of irradiance (representing intensity value of pixels) is blurred by the acquisition device. However, in practice it is reasonable and realistic to assumed that the blur imaging device is rather moderated ; hence local areas around edges are not orthogonal to the plane $z=0$ but tend to have the highest variation or gradient. 

As we don't know the analytic expression of the surface, the direct use of Taylor approximation coefficients is impossible. The coefficients have to be estimated. There are several way to locally approximate a surface \cite{LASIP06}. Several methods based on pruning \cite{Shukla05IP,Kazinnik07IP} could accelerate the proposed method but for a more accurate detector, it should be better to work on small regions of interest. Thus, the least square method is used.

The RSD method proposes to approximate an area of the image by using a regression surface based on the image content model detailed in Section~\ref{sec:image_model}. To this end, the regression's coefficients are calculated for each pixel's neighborhood in the image. Those coefficients are analyzed in order to determine if that neighborhood contains an edge using Equation~(\ref{eq:edge_Taylor}) or an equivalent through an other axe in the image plan. This approach can also be used to determine what kind of edge it is because edge orientation and curvature have an high influence on Taylor approximation's coefficients.

In the rest of this paper, the $c_k$ are the regression's coefficients. The description of proposed methodology focuses on grayscale images because, as previously mentioned, each color channel is analyzed separately in the case of color images. 

\subsection{Surface Regression}\label{sec:surface_regression}
A set of $N^2$ equations have to be defined using the intensity value $I(x_0+\bar{x}_n,y_0+\bar{y}_n)$ of each pixel of the relative coordinates $(\bar{x}_n,\bar{y}_n)$ in the neighborhood $\V_{x_0,y_0}$ in order to regress the surface to a $K$-degree polynomial. Such a model has, for instance, shown its efficiency in hidden information to pick the most relevant area for embedding~\cite{VS2016TIFS,VS2015SPIE}. This means that we have to solve the following overconstrained system:\\
\begin{equation}\label{eq:system}
	A_K C_K = B    
\end{equation}
where $A_K$ is the matrix of size $N^2 \times (K+\frac{K(K+1)}{2})$ encoding the relative coordinates for a bi-variate polynomial regression surface of order $K$, $B$ is the column vector of $N^2$ components representing pixels' intensity values in the lexical order and $C_K$ represents the polynomial coefficients. 
Since the system~(\ref{eq:system}) is overconstraint, no solution might exists; therefore it is proposed to solve it with the least-square method:
\begin{equation}\label{eq:systemLS}
    \mathrm{min} \lVert A_K C_K - B \rVert_2^2
\end{equation}
A quadric for example, leads to this system:
\begin{equation*}
	\mathrm{min} \lVert A_2 C_2 - B \rVert_2^2
\end{equation*}
with
\begin{equation}\label{eq:def_matrix_A_reg}
	A_2=
	\begin{pmatrix}
		1 & \bar{x}_1 & \bar{y}_1 & \bar{x}_1^2 & 			\bar{y}_1^2 & \bar{x}_1 \bar{y}_1\\
		1 & \bar{x}_1 & \bar{y}_2 & \bar{x}_1^2 & 			\bar{y}_2^2 & \bar{x}_1 \bar{y}_2\\
		& \vdots & & \vdots & & \vdots \\
		1 & \bar{x}_N & \bar{y}_{N-1} & \bar{x}_N^2 & \bar{y}_{N-1}^2 & 				\bar{x}_N \bar{y}_{N-1}\\
		1 & \bar{x}_N & \bar{y}_N & \bar{x}_N^2 & \bar{y}_N^2 & 				\bar{x}_N \bar{y}_N
	\end{pmatrix}
\end{equation}
\begin{equation*}	
	C_2=\begin{pmatrix} c_0 & c_1 & c_2 & c_3 & c_4 & c_5 				\end{pmatrix}^T
\end{equation*}
\begin{equation*}	
	B=\begin{pmatrix} I(x_0+\bar{x}_1,y_0+\bar{y}_1) &  \cdots  & 		I(x_0+\bar{x}_N,y_0+\bar{y}_N) \end{pmatrix}^T
\end{equation*}

The matrix $A_K$ is independent of $(x_0,y_0)$ due to the relative coordinates. Thus $A_K^T A_K$ is also constant and can not be affected by noise. If it is a nonsingular matrix, Equation~(\ref{eq:systemLS}) has a unique solution which can be found by solving the equivalent problem:
\begin{equation}\label{eq:least_square_system}
    \mathrm{min} \lVert A_K^T A_K C_K - A_K^T B \rVert_2^2
\end{equation}

In fact, $A_K$ is full rank column as the base of two variable's polynomial could be retrieved in rows. Hence, the solution of the Equation~(\ref{eq:systemLS}) is unique and $A_K^T A_K$ is inversible. 
The calculations details lead, from Gauss-Markov theorem and Equation~(\ref{eq:least_square_system}), to the well-known Best Linear Unbiased Estimator (BLUE \cite{BLUE1973}) of coefficients $C_K$:
\begin{equation}\label{eq:least_square_estimation}
	C_K = \left( A_K^T A_K \right)^{-1} A_K^T B
\end{equation}

But unfortunately, it is also ill-conditioned ($Cond_2(A_3^T A_3)>600$ for $N=3$). This can lead to numerically unstable approximation as the condition number represents the potential relative estimation error for a given error on measure (pixel intensity in this case): 
\begin{equation*}
	\frac{\|\Delta C_K \|}{\|C_K\|} \leq Cond(A_K^T A_K) \frac{\|			A_K^T \Delta B\|}{\|A_K^T B\|}
\end{equation*}  

Compared to the direct matrix inverse, inverse solutions using QR decomposition are more numerically stable as evidenced by their reduced condition numbers \cite[chap1.13]{parker1994geophysical}. Thus, in order to deal with this stability problem, a Modified Gram-Schmidt\cite[chap5.2.8]{loan1996matrix} QR factorization is used: $A_K = Q_K \left[ \begin{matrix} R_K^T & \text{\O} \end{matrix} \right]^T $ with \text{\O} a null matrix used to have the good number of row. Note that $R_K^T$ is a square matrix. 
The solution~(\ref{eq:least_square_estimation}) could be reformulated using the orthogonality of $Q$:
\begin{equation}
	C_K = \left[\begin{matrix}R_K^{-1}&\text{\O}\end{matrix}\right] Q_K^T B
\end{equation}

Thus 
$\left[\begin{matrix}R_K^{-1}&\text{\O}\end{matrix}\right] Q_K^T$ 
only has to be calculated once in order to find regression's coefficients in the neighborhood $\V_{x_0,y_0}$ of each pixel $(x_0,y_0)$ in the image.

\subsection{Edge detection}\label{sec:edge_detection}

\begin{figure*}
\centering
\includegraphics[width=0.235\textwidth, height=0.235\textwidth, trim=0pt 125pt 200pt 75pt , clip]{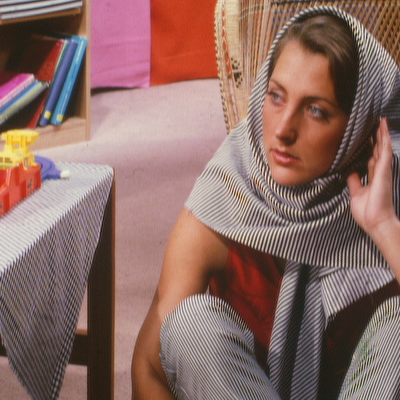}
\\
\renewcommand{\arraystretch}{0}
\begin{tabular}{lm{0.235\textwidth}@{}*{4}{@{}m{0.235\textwidth}}}
&\begin{center}$3\times3$\end{center}
&\begin{center}$5\times5$\end{center}
&\begin{center}$7\times7$\end{center}
&\begin{center}$9\times9$\end{center}
\\1
&\includegraphics[width=0.235\textwidth, height=0.235\textwidth, trim=0pt 125pt 200pt 75pt , clip]{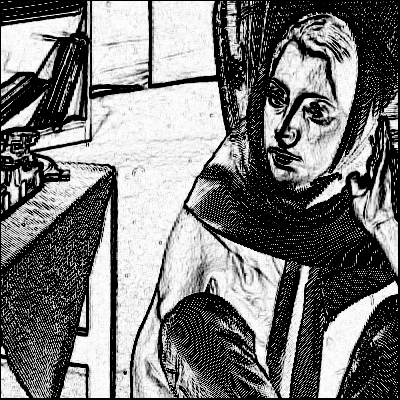}
&\includegraphics[width=0.235\textwidth, height=0.235\textwidth, trim=0pt 125pt 200pt 75pt , clip]{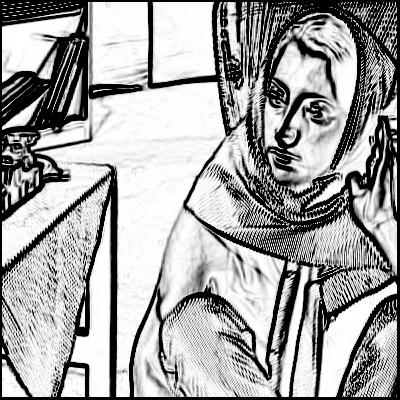}
&\includegraphics[width=0.235\textwidth, height=0.235\textwidth, trim=0pt 125pt 200pt 75pt , clip]{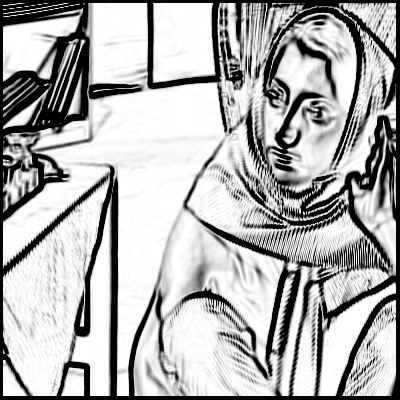}
&\includegraphics[width=0.235\textwidth, height=0.235\textwidth, trim=0pt 125pt 200pt 75pt , clip]{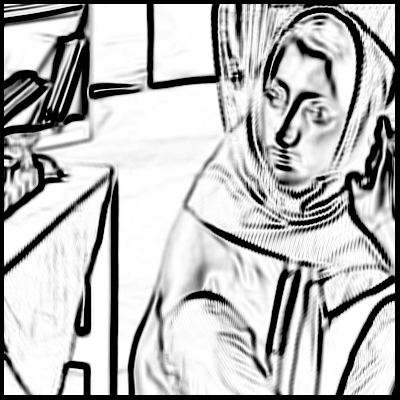}
\\2
&\includegraphics[width=0.235\textwidth, height=0.235\textwidth, trim=0pt 125pt 200pt 75pt , clip]{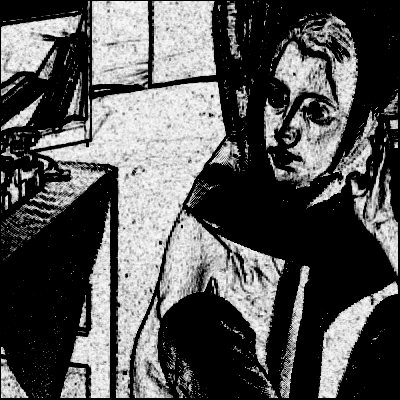}
&\includegraphics[width=0.235\textwidth, height=0.235\textwidth, trim=0pt 125pt 200pt 75pt , clip]{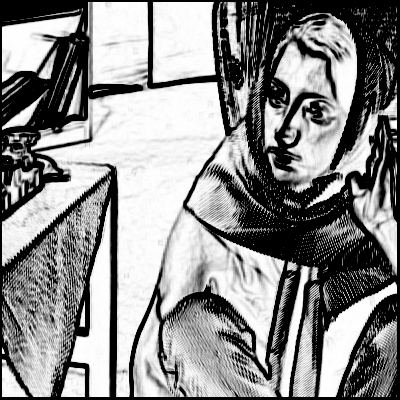}
&\includegraphics[width=0.235\textwidth, height=0.235\textwidth, trim=0pt 125pt 200pt 75pt , clip]{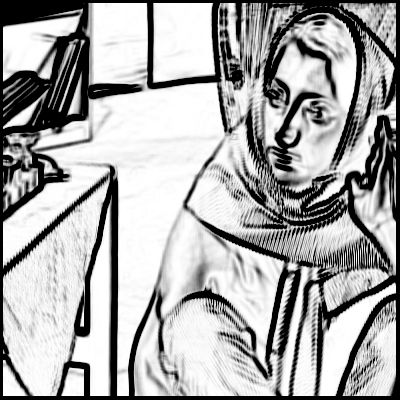}
&\includegraphics[width=0.235\textwidth, height=0.235\textwidth, trim=0pt 125pt 200pt 75pt , clip]{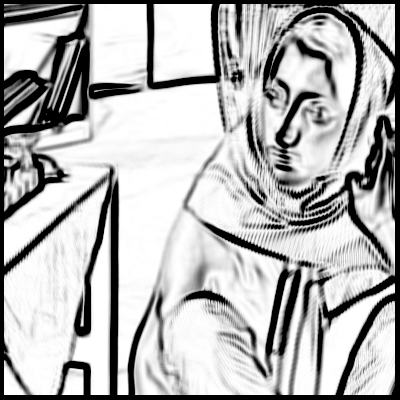}
\\3
&&\includegraphics[width=0.235\textwidth, height=0.235\textwidth, trim=0pt 125pt 200pt 75pt , clip]{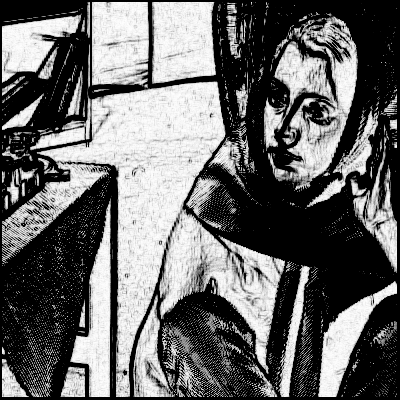}
&\includegraphics[width=0.235\textwidth, height=0.235\textwidth, trim=0pt 125pt 200pt 75pt , clip]{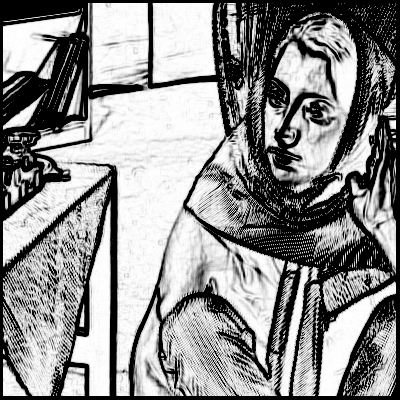}
&\includegraphics[width=0.235\textwidth, height=0.235\textwidth, trim=0pt 125pt 200pt 75pt , clip]{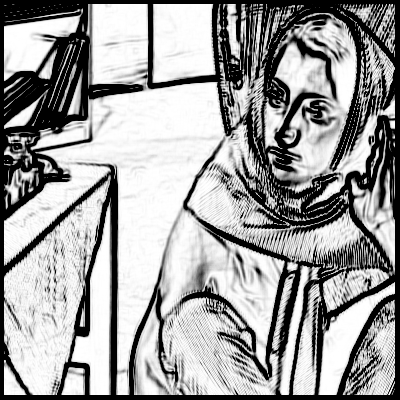}
\\4
&&\includegraphics[width=0.235\textwidth, height=0.235\textwidth, trim=0pt 125pt 200pt 75pt , clip]{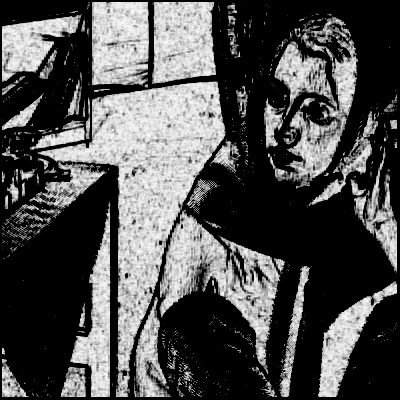}
&\includegraphics[width=0.235\textwidth, height=0.235\textwidth, trim=0pt 125pt 200pt 75pt , clip]{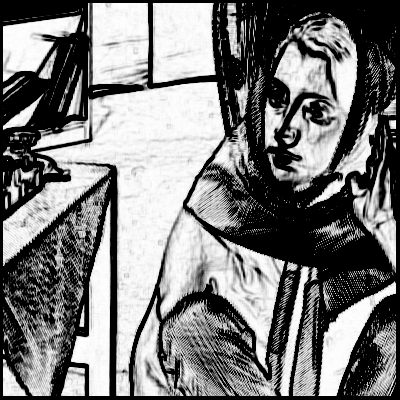}
&\includegraphics[width=0.235\textwidth, height=0.235\textwidth, trim=0pt 125pt 200pt 75pt , clip]{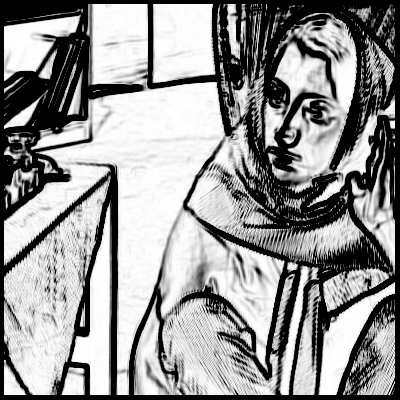}
\end{tabular}
\caption{Edges map resulting of \emph{Barbara} image. Rows correspond to the order of the regression surface and columns correspond to the neighborhood area used. Note that for display purposes, the highest norm values, which are expected to be edges, are shown in black.}
\label{fig:barbara_edgestest}
\end{figure*}

As it is explained previously with the Taylor expansion~(\ref{eq:edge_Taylor}) of the image content model given in Section~\ref{subsec:image_model_image}, regression's coefficients are a good indicator for edges presence. Each coefficients have a different sensitivity to different edges shape. As it have been demonstrated with Equation~(\ref{eq:edge_Taylor}) and illustrated with Figure~\ref{fig:testimagec1}, the $c_1$ coefficient reacts with edges of vertical orientation. 

All orientations and shapes are a different combination of all the coefficients $c_k$ except $c_0$ which corresponds to the $0$ order element of the regression. In the proposed model, this coefficient could be seen as the influence of the continuous part of Equation~({\ref{eq:model_scene_decomp}) which appears as $I_c$ in Equation~({\ref{eq:model_image_decomp}). Because of this, the $c_0$ coefficient can not be used. All other coefficients tend to have a significant value in presence of an edge. See Section~\ref{sec:edge_description} for a thorough analysis of edge's reponse.

The easiest way to compute an edge map knowing this is to calculate the norm of the coefficients $c_k$, with $k>0$. Euclidean norm is used in our implementation. The result could be easily segmented with a threshold.

For an edge map in a color image, it is possible to calculate the norm of the regression's coefficients (without $c_0$) in each concatenated channel. This leads to direct fusion of each channel result, see Figure~\ref{fig:barbara_edgestest} for a color example with different parameters.

As expected, using the RSD method on a large area leads to remove high frequency variation in the edge detection. This is due to the mean implicitly done by the least-square method in order to find the minimum solution of system~(\ref{eq:system}). This could be used in order to remove unwanted edges detected in textured area as we could see in Figure~\ref{fig:barbara_edgestest}.

On the contrary, increasing the order of the surface for the regression leads to detect more details in the edge map. This is due to the capacity of higher order polynomial to better fit the shape of the image discrete surface. But those two parameters are not bound and we could use this fact. For example, the edge map calculated on \emph{Barbara}, see Figure~\ref{fig:barbara_edgestest} with a neighborhood area of $3\times3$ regressed with planar surface leads to an edge map with important noise in the textured area unlike the one calculated with a neighborhood of $9\times9$ regressed on a fourth order surface. But in all case, most important edges, which correspond to objects boundaries, tend to be preserved.

Note that for display purpose, we have inverted the color map with this function for all the results in Figures~\ref{fig:barbara_edgestest} - \ref{fig:edge_recognition_boats}:
\begin{equation*}
	\begin{cases}
		255-\lfloor Norm^2\rfloor &\text{when}~~Norm^2 <255\\
		0&\text{otherwise}
	\end{cases}
\end{equation*}

\subsection{Edge and corner descriptor}\label{sec:edge_description}

\begin{figure}[!t]
\centering
\begin{tabular}{@{}cc@{}}
\subfloat[A test image in color.]{
  \includegraphics[width=0.22\textwidth , trim=5 6 6 5 , clip]{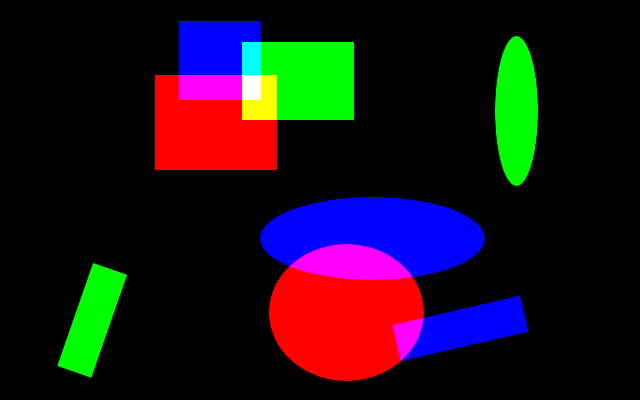}
  \label{fig:colortest} }
&
\subfloat[Shapes S1-S4.]{
  \includegraphics[width=0.22\textwidth , height=1cm]{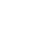}
  \put(-80,30){\includegraphics[width=0.05\textwidth]{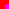}}
  \put(-50,30){\includegraphics[width=0.05\textwidth]{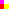}}
  \put(-80,0){\includegraphics[width=0.05\textwidth]{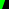}}
  \put(-50,0){\includegraphics[width=0.05\textwidth]{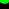}}
\label{fig:boats_color_pattern} }
\\
  \subfloat[Similar edges founded with pattern P1 on test.]{
  \setlength\fboxsep{0pt}
  \fbox{\includegraphics[width=0.22\textwidth , trim=5 6 6 5 , clip]{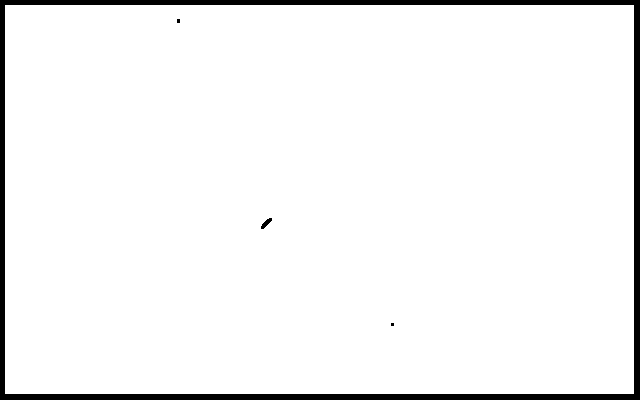}}
  \label{fig:colortest_edge_descriptor_P1} }
  &
  \subfloat[Similar edges founded with pattern P2 on test.]{
  \setlength\fboxsep{0pt}
  \fbox{\includegraphics[width=0.22\textwidth , trim=5 6 6 5 , clip]{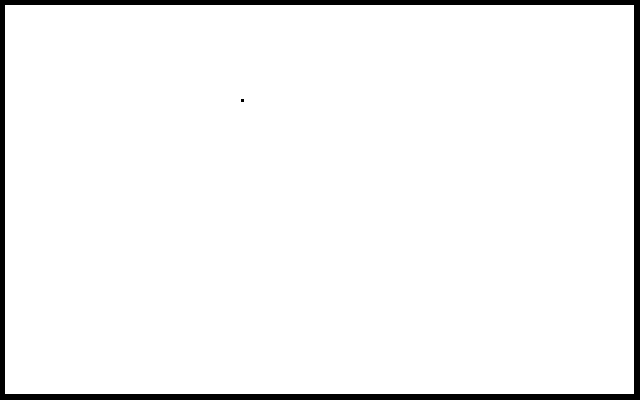}}
  \label{fig:colortest_edge_descriptor_P2} }
  \\
  \subfloat[Similar edges founded with pattern P3 on test.]{
  \setlength\fboxsep{0pt}
  \fbox{\includegraphics[width=0.22\textwidth , trim=5 6 6 5 , clip]{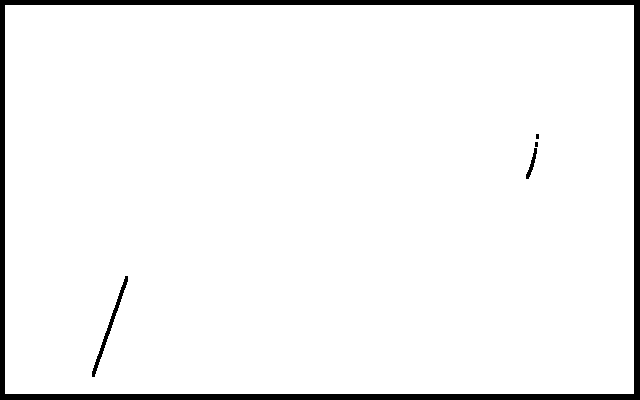}}
  \label{fig:colortest_edge_descriptor_P3} }  
  &
  \subfloat[Similar edges founded with pattern P4 on test.]{
  \setlength\fboxsep{0pt}
  \fbox{\includegraphics[width=0.22\textwidth , trim=5 6 6 5 , clip]{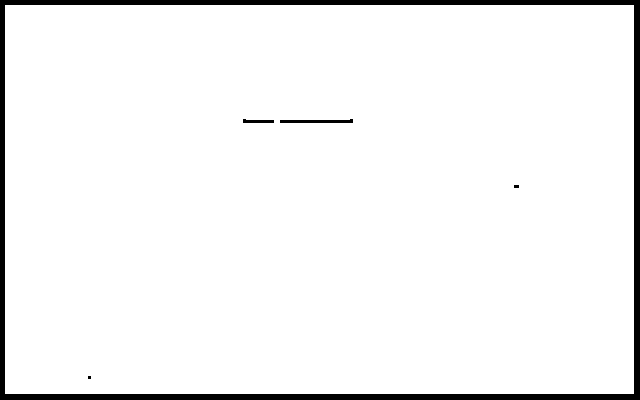}}
  \label{fig:colortest_edge_descriptor_P4} }
\end{tabular}
\caption{Results of the proposed edge pattern recognition. The blocks nearest to each patter are shown in black. Note that for display purpose the detected pattern are enlarged by morphological dilation.}
\label{fig:edge_recognition_colortest}
\end{figure}

\begin{figure}[!t]
\centering
\begin{tabular}{cc}
\subfloat[\emph{Boats} image.]{
  \includegraphics[width=0.22\textwidth , trim=5 6 6 5 , clip]{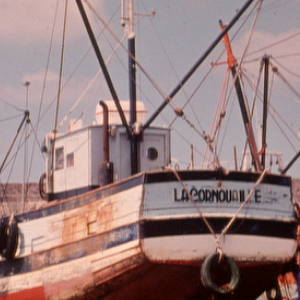}
  \label{fig:boats} }
&
  \subfloat[Patterns P5-P8.]{
  \includegraphics[width=0.22\textwidth , height=1cm]{blank.png}
  \put(-80,30){\includegraphics[width=0.05\textwidth]{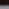}}
  \put(-50,30){\includegraphics[width=0.05\textwidth]{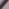}}
  \put(-80,0){\includegraphics[width=0.05\textwidth]{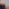}}
  \put(-50,0){\includegraphics[width=0.05\textwidth]{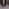}}
  \label{fig:boats_edge_pattern} }
\\
  \subfloat[Similar edges founded with pattern P5 on boats.]{
  \setlength\fboxsep{0pt}
  \fbox{\includegraphics[width=0.22\textwidth , trim=5 6 6 5 , clip]{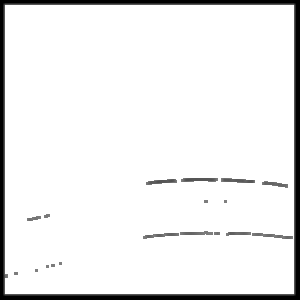}}
  \label{fig:boats_edge_descriptor_P5} }
  &
  \subfloat[Similar edges founded with pattern P6 on boats.]{
  \setlength\fboxsep{0pt}
  \fbox{\includegraphics[width=0.22\textwidth , trim=5 6 6 5 , clip]{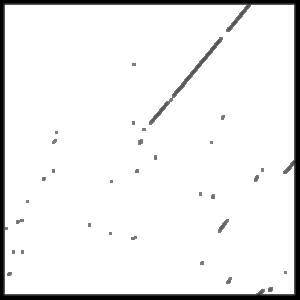}}
  \label{fig:boats_edge_descriptor_P6} }
  \\
  \subfloat[Similar edges founded with pattern P7 on boats.]{
  \setlength\fboxsep{0pt}
  \fbox{\includegraphics[width=0.22\textwidth , trim=5 6 6 5 , clip]{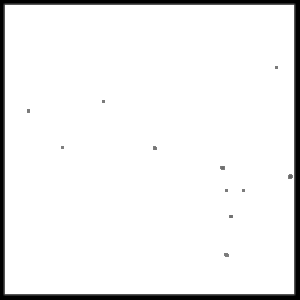}}
  \label{fig:boats_edge_descriptor_P7} }  
  &
  \subfloat[Similar edges founded with pattern P8 on boats.]{
  \setlength\fboxsep{0pt}
  \fbox{\includegraphics[width=0.22\textwidth , trim=5 6 6 5 , clip]{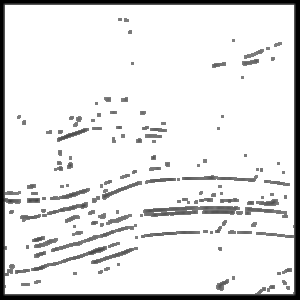}}
  \label{fig:boats_edge_descriptor_P8} }
\end{tabular}
\caption{Results of the proposed edge pattern recognition. The blocks nearest to each patter are shown in black. Note that for display purpose the detected pattern are enlarged by morphological dilation.}
\label{fig:edge_recognition_boats}
\end{figure}

The regression's coefficients can be analyzed in order to determine in which condition they could be non zero. Coefficients react differently depending of the edge direction and curvature. Thus, it is possible to describe any point on an edge with them.

It is also possible to search all edges similar to a defined pattern using a distance between the coefficients; formally, for retrieving edges similar to a reference which gives coefficients $c^\star_K$ this corresponds to compute the quantity $\lVert c^\star_K - c_K \rVert$. See Figure~\ref{fig:edge_recognition_colortest} and \ref{fig:edge_recognition_boats} for examples of results obtained with this methodology. In our test, the Euclidean distance is used.

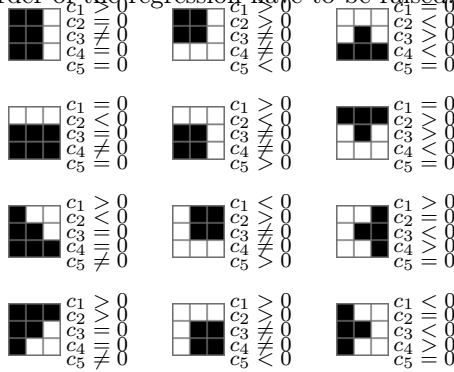
\begin{figure}[b!]
  \centering
  \def\svgwidth{0.47\textwidth}
\begingroup
  \makeatletter
    \setlength{\unitlength}{\svgwidth}
  \makeatother
\begin{picture}(0.91,0.6)%
    \put(0,0.6){\includegraphics[width=0.05\textwidth]{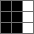}}%
    \put(0,0.4){\includegraphics[width=0.05\textwidth]{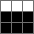}}%
    \put(0,0.2){\includegraphics[width=0.05\textwidth]{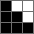}}%
    \put(0,0){\includegraphics[width=0.05\textwidth]{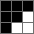}}%
    \put(0.333,0.6){\includegraphics[width=0.05\textwidth]{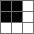}}%
    \put(0.333,0.4){\includegraphics[width=0.05\textwidth]{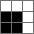}}%
    \put(0.333,0.2){\includegraphics[width=0.05\textwidth]{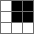}}%
    \put(0.333,0){\includegraphics[width=0.05\textwidth]{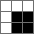}}%
    \put(0.666,0.6){\includegraphics[width=0.05\textwidth]{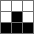}}%
    \put(0.666,0.4){\includegraphics[width=0.05\textwidth]{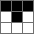}}%
    \put(0.666,0.2){\includegraphics[width=0.05\textwidth]{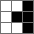}}%
    \put(0.666,0){\includegraphics[width=0.05\textwidth]{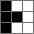}}%
\begin{footnotesize}
    \put(0.18,0.7){\color[rgb]{0,0,0}\makebox(0,0)[b]{\smash{$c_1>0$}}}%
    \put(0.18,0.67){\color[rgb]{0,0,0}\makebox(0,0)[b]{\smash{$c_2=0$}}}%
    \put(0.18,0.64){\color[rgb]{0,0,0}\makebox(0,0)[b]{\smash{$c_3\neq0$}}}%
    \put(0.18,0.61){\color[rgb]{0,0,0}\makebox(0,0)[b]{\smash{$c_4=0$}}}%
    \put(0.18,0.58){\color[rgb]{0,0,0}\makebox(0,0)[b]{\smash{$c_5=0$}}}%
    \put(0.18,0.5){\color[rgb]{0,0,0}\makebox(0,0)[b]{\smash{$c_1=0$}}}%
    \put(0.18,0.47){\color[rgb]{0,0,0}\makebox(0,0)[b]{\smash{$c_2<0$}}}%
    \put(0.18,0.44){\color[rgb]{0,0,0}\makebox(0,0)[b]{\smash{$c_3=0$}}}%
    \put(0.18,0.41){\color[rgb]{0,0,0}\makebox(0,0)[b]{\smash{$c_4\neq0$}}}%
    \put(0.18,0.38){\color[rgb]{0,0,0}\makebox(0,0)[b]{\smash{$c_5=0$}}}%
    \put(0.18,0.3){\color[rgb]{0,0,0}\makebox(0,0)[b]{\smash{$c_1>0$}}}%
    \put(0.18,0.27){\color[rgb]{0,0,0}\makebox(0,0)[b]{\smash{$c_2<0$}}}%
    \put(0.18,0.24){\color[rgb]{0,0,0}\makebox(0,0)[b]{\smash{$c_3=0$}}}%
    \put(0.18,0.21){\color[rgb]{0,0,0}\makebox(0,0)[b]{\smash{$c_4=0$}}}%
    \put(0.18,0.18){\color[rgb]{0,0,0}\makebox(0,0)[b]{\smash{$c_5\neq0$}}}%
    \put(0.18,0.1){\color[rgb]{0,0,0}\makebox(0,0)[b]{\smash{$c_1>0$}}}%
    \put(0.18,0.07){\color[rgb]{0,0,0}\makebox(0,0)[b]{\smash{$c_2>0$}}}%
    \put(0.18,0.04){\color[rgb]{0,0,0}\makebox(0,0)[b]{\smash{$c_3=0$}}}%
    \put(0.18,0.01){\color[rgb]{0,0,0}\makebox(0,0)[b]{\smash{$c_4=0$}}}%
    \put(0.18,-0.02){\color[rgb]{0,0,0}\makebox(0,0)[b]{\smash{$c_5\neq0$}}}%
    \put(0.513,0.7){\color[rgb]{0,0,0}\makebox(0,0)[b]{\smash{$c_1>0$}}}%
    \put(0.513,0.67){\color[rgb]{0,0,0}\makebox(0,0)[b]{\smash{$c_2>0$}}}%
    \put(0.513,0.64){\color[rgb]{0,0,0}\makebox(0,0)[b]{\smash{$c_3\neq0$}}}%
    \put(0.513,0.61){\color[rgb]{0,0,0}\makebox(0,0)[b]{\smash{$c_4\neq0$}}}%
    \put(0.513,0.58){\color[rgb]{0,0,0}\makebox(0,0)[b]{\smash{$c_5<0$}}}%
    \put(0.513,0.5){\color[rgb]{0,0,0}\makebox(0,0)[b]{\smash{$c_1>0$}}}%
    \put(0.513,0.47){\color[rgb]{0,0,0}\makebox(0,0)[b]{\smash{$c_2<0$}}}%
    \put(0.513,0.44){\color[rgb]{0,0,0}\makebox(0,0)[b]{\smash{$c_3\neq0$}}}%
    \put(0.513,0.41){\color[rgb]{0,0,0}\makebox(0,0)[b]{\smash{$c_4\neq0$}}}%
    \put(0.513,0.38){\color[rgb]{0,0,0}\makebox(0,0)[b]{\smash{$c_5>0$}}}%
    \put(0.513,0.3){\color[rgb]{0,0,0}\makebox(0,0)[b]{\smash{$c_1<0$}}}%
    \put(0.513,0.27){\color[rgb]{0,0,0}\makebox(0,0)[b]{\smash{$c_2>0$}}}%
    \put(0.513,0.24){\color[rgb]{0,0,0}\makebox(0,0)[b]{\smash{$c_3\neq0$}}}%
    \put(0.513,0.21){\color[rgb]{0,0,0}\makebox(0,0)[b]{\smash{$c_4\neq0$}}}%
    \put(0.513,0.18){\color[rgb]{0,0,0}\makebox(0,0)[b]{\smash{$c_5>0$}}}%
    \put(0.513,0.1){\color[rgb]{0,0,0}\makebox(0,0)[b]{\smash{$c_1>0$}}}%
    \put(0.513,0.07){\color[rgb]{0,0,0}\makebox(0,0)[b]{\smash{$c_2>0$}}}%
    \put(0.513,0.04){\color[rgb]{0,0,0}\makebox(0,0)[b]{\smash{$c_3\neq0$}}}%
    \put(0.513,0.01){\color[rgb]{0,0,0}\makebox(0,0)[b]{\smash{$c_4\neq0$}}}%
    \put(0.513,-0.02){\color[rgb]{0,0,0}\makebox(0,0)[b]{\smash{$c_5<0$}}}%
    \put(0.846,0.7){\color[rgb]{0,0,0}\makebox(0,0)[b]{\smash{$c_1=0$}}}%
    \put(0.846,0.67){\color[rgb]{0,0,0}\makebox(0,0)[b]{\smash{$c_2<0$}}}%
    \put(0.846,0.64){\color[rgb]{0,0,0}\makebox(0,0)[b]{\smash{$c_3>0$}}}%
    \put(0.846,0.61){\color[rgb]{0,0,0}\makebox(0,0)[b]{\smash{$c_4<0$}}}%
    \put(0.846,0.58){\color[rgb]{0,0,0}\makebox(0,0)[b]{\smash{$c_5=0$}}}%
    \put(0.846,0.5){\color[rgb]{0,0,0}\makebox(0,0)[b]{\smash{$c_1=0$}}}%
    \put(0.846,0.47){\color[rgb]{0,0,0}\makebox(0,0)[b]{\smash{$c_2>0$}}}%
    \put(0.846,0.44){\color[rgb]{0,0,0}\makebox(0,0)[b]{\smash{$c_3>0$}}}%
    \put(0.846,0.41){\color[rgb]{0,0,0}\makebox(0,0)[b]{\smash{$c_4<0$}}}%
    \put(0.846,0.38){\color[rgb]{0,0,0}\makebox(0,0)[b]{\smash{$c_5=0$}}}%
    \put(0.846,0.3){\color[rgb]{0,0,0}\makebox(0,0)[b]{\smash{$c_1>0$}}}%
    \put(0.846,0.27){\color[rgb]{0,0,0}\makebox(0,0)[b]{\smash{$c_2=0$}}}%
    \put(0.846,0.24){\color[rgb]{0,0,0}\makebox(0,0)[b]{\smash{$c_3<0$}}}%
    \put(0.846,0.21){\color[rgb]{0,0,0}\makebox(0,0)[b]{\smash{$c_4>0$}}}%
    \put(0.846,0.18){\color[rgb]{0,0,0}\makebox(0,0)[b]{\smash{$c_5=0$}}}%
    \put(0.846,0.1){\color[rgb]{0,0,0}\makebox(0,0)[b]{\smash{$c_1<0$}}}%
    \put(0.846,0.07){\color[rgb]{0,0,0}\makebox(0,0)[b]{\smash{$c_2=0$}}}%
    \put(0.846,0.04){\color[rgb]{0,0,0}\makebox(0,0)[b]{\smash{$c_3<0$}}}%
    \put(0.846,0.01){\color[rgb]{0,0,0}\makebox(0,0)[b]{\smash{$c_4>0$}}}%
    \put(0.846,-0.02){\color[rgb]{0,0,0}\makebox(0,0)[b]{\smash{$c_5=0$}}}%
\end{footnotesize}
\end{picture}
\endgroup
\caption{Coeficients analysis on case exemples.}
\label{fig:coefficients_analysis}
\end{figure}

Again, the $c_0$ coefficient is not used in the calculation of distance. Other coefficients respond differently in presence of discontinuity. Figure~\ref{fig:coefficients_analysis} shows coefficients' response in several simple cases. Only the sign and zero value have to be considered in order to retrieve those shapes. The value of each coefficient depends of the intensity difference around the edge. Those results stay the same if the pattern is shifted by one pixel. That is why the symbol ``$\neq$'' is employed (the sign change on certain shifts). Those results can be obtained by analysing the Taylor coefficients or by considering directly the shape of the regression.

Empirically, we have found that complex edges arrangement could lead to many false detections, especially if the order of the regressing surface is to small. For the given example, we have used a $9\times9$ neighborhood in order to have a significant area for the comparison.

Note that Harris \cite{HarrisCorner} and Shi-Tomasi \cite{ShiTomasiCorner} corner detectors are computed using, as in the proposed method, a Taylor expansion of the squared differences. The Harris matrix can be retrieved in $A_K^T A_K$ on the diagonal 
when the least square solution is computed~(\ref{eq:least_square_system}). The coefficients $c_3$, $c_4$ and $c_5$ are calculated using this part of $A_K^T A_K$.

In practice, we could see that $c_3$, $c_4$ and $c_5$ highly react in presence of a corner in the considered neighborhood. Some example of corner retrieval could be seen in Figure~\ref{fig:colortest_edge_descriptor_P1}, \ref{fig:colortest_edge_descriptor_P2} and \ref{fig:boats_edge_descriptor_P7}.

The proposed model can be used in order to find edges of a certain curvature. This works almost perfectly for rather simple images, see Figure~\ref{fig:colortest_edge_descriptor_P4} but in practice with a real image, noise has a large influence on obtained results as shown in \ref{fig:boats_edge_descriptor_P8}. In order to deal with complex curvature, the order of the regression have to be raised.

\section{Numerical Results and Comparison with Other Detectors}\label{sec:num_results}

Before the numerical results' presentation, it is proposed to briefly explain the robustness of the proposed method.

In order to test the efficiency of the proposed model and detection methods, we have constructed a sample on which is calculated the signal noise ratio in different cases of image's alteration like Gaussian noise or blur.

A comparison with other classical algorithms in edges detection for the same sample is also exposed below.

\subsection{Robustness to Blurring Process and Additive Noise}\label{sec:num_res_blur_noise}

%

\begin{small}
\begin{table*}[ht!]
\begin{center}
\hspace*{-3.5cm}
  \begin{tabular}{l|c|c|c|c|c}
  Detector	\!\!&\!\!	blur rad. $s\!=\!1$	\!\!&\!\!	blur rad. $s\!=\!2$	\!\!&\!\!	blur rad. $s\!=\!4$	\!\!&\!\!	blur rad. $s\!=\!7$	\!\!&\!\!	blur rad. $s\!=\!10$\\
					\!\!&\!\!	PSNR $\!=\!25.80$		\!\!&\!\!	PSNR $\!=\!21.52$		\!\!&\!\!	PSNR $\!=\!18.83$		\!\!&\!\!	PSNR $\!=\!16.41$		\!\!&\!\!	PSNR $\!=\!14.81$ \\
  \hline
	Canny (scale \!=\!1)  			\!\!&\!\!	$P_{MD}\!=\!40.36\%$	\!\!&\!\!	$P_{MD}\!=\!59.55\%$	\!\!&\!\!	$P_{MD}\!=\!66.90\%$	\!\!&\!\!	$P_{MD}\!=\!78.37\%$	\!\!&\!\!	$P_{MD}\!=\!81.95\%$\\
	Canny (scale \!=\!2)  			\!\!&\!\!	$P_{MD}\!=\!36.44\%$	\!\!&\!\!	$P_{MD}\!=\!60.81\%$	\!\!&\!\!	$P_{MD}\!=\!70.79\%$	\!\!&\!\!	$P_{MD}\!=\!86.37\%$	\!\!&\!\!	$P_{MD}\!=\!89.04\%$ \\
	LoG (scale \!=\!1) 		\!\!&\!\!	$P_{MD}\!=\!12.94\%$	\!\!&\!\!	$P_{MD}\!=\!51.09\%$	\!\!&\!\!	$P_{MD}\!=\!57.82\%$	\!\!&\!\!	$P_{MD}\!=\!\mathbf{\textcolor{blue}{56.40\%}}$	\!\!&\!\!	$P_{MD}\!=\!\mathbf{\textcolor{blue}{54.90}}\%$\\
	LoG (scale \!=\!2) 		\!\!&\!\!	$P_{MD}\!=\!{24.12\%}$	\!\!&\!\!	$P_{MD}\!=\!64.91\%$	\!\!&\!\!	$P_{MD}\!=\!77.06\%$	\!\!&\!\!	$P_{MD}\!=\!80.35\%$	\!\!&\!\!	$P_{MD}\!=\!84.02\%$ \\
	NL-filter~\cite{Laligant2010}	\!\!&\!\!	$P_{MD}\!=\!19.61\%$	\!\!&\!\!	$P_{MD}\!=\!43.65\%$	\!\!&\!\!	$P_{MD}\!=\!56.86\%$	\!\!&\!\!	$P_{MD}\!=\!{74.78\%}$	\!\!&\!\!	$P_{MD}\!=\!{81.05\%}$ \\
	NL-filter~\cite{Laligant2010} with median filter 	\!\!&\!\!	$P_{MD}\!=\!4.26\%$	\!\!&\!\!	$P_{MD}\!=\!24.85\%$	\!\!&\!\!	$P_{MD}\!=\!\mathbf{\textcolor{blue}{43.11\%}}$	\!\!&\!\!	$P_{MD}\!=\!56.52\%$	\!\!&\!\!	$P_{MD}\!=\!63.24\%$\\
	Haralick Facet~\cite{Haralick84} \!\!&\!\!	$P_{MD}\!=\!54.68\%$	\!\!&\!\!	$P_{MD}\!=\!73.82\%$	\!\!&\!\!	$P_{MD}\!=\!80.44\%$	\!\!&\!\!	$P_{MD}\!=\!98.80\%$	\!\!&\!\!	$P_{MD}\!=\!99.95\%$\\
	Haralick Facet~\cite{Haralick84} with median filter	\!\!&\!\!	$P_{MD}\!=\!\emph{\textcolor{red}{55.52\%}}$	\!\!&\!\!	$P_{MD}\!=\!\emph{\textcolor{red}{75.29\%}}$	\!\!&\!\!	$P_{MD}\!=\!\emph{\textcolor{red}{80.58\%}}$	\!\!&\!\!	$P_{MD}\!=\!\emph{\textcolor{red}{98.85\%}}$	\!\!&\!\!	$P_{MD}\!=\!\emph{\textcolor{red}{99.96\%}}$\\
	Proposed method	$\alpha$ ($3\times3$, order 2)\!\!&\!\!	$P_{MD}\!=\!\mathbf{\textcolor{blue}{0.05\%}}$	\!\!&\!\!	$P_{MD}\!=\!\mathbf{\textcolor{blue}{21.39\%}}$	\!\!&\!\!	$P_{MD}\!=\!46.42\%$	\!\!&\!\!	$P_{MD}\!=\!58.61\%$	\!\!&\!\!	$P_{MD}\!=\!68.88\%$ \\
	Proposed method	$\beta$ ($5\times5$, order 2)\!\!&\!\!	$P_{MD}\!=\!5.15\%$	\!\!&\!\!	$P_{MD}\!=\!28.44\%$	\!\!&\!\!	$P_{MD}\!=\!46.46\%$	\!\!&\!\!	$P_{MD}\!=\!58.76\%$	\!\!&\!\!	$P_{MD}\!=\!65.12\%$ 
  \end{tabular}
\end{center}
\caption{Numerical comparison of edge detectors robustness to invariant Gaussian blur ; best results are shown in bold blue and worst results are shown are displayed in italic red.}
\label{tbl:robustness_blurred_noise}
\end{table*}
\end{small}


\begin{figure}[b!]
\centerline{
\hspace*{-0.15cm}
  \subfloat[Test image used for robustness evaluation.]{
\hspace*{-0.1cm}
	\includegraphics[width=0.22\textwidth, height=0.20\textwidth]{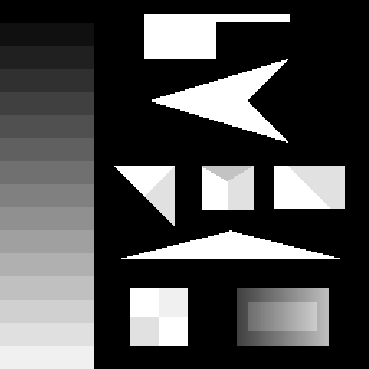}
  \label{fig:imtest_Mire}	}
\hspace{0.005\columnwidth}
\hspace*{0.1cm}
  \subfloat[Ground truth used as edge location in the test image.]{
\hspace*{-0.1cm}
	\includegraphics[width=0.22\textwidth, height=0.20\textwidth]{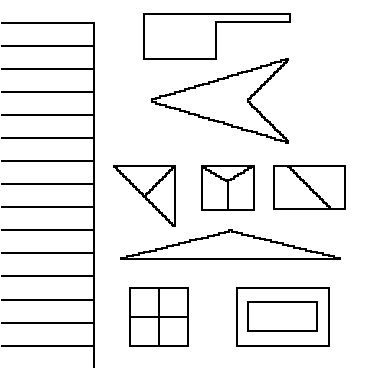}
  \label{fig:imtest_ground_truth}	}	}
\caption{Digital image used for evaluating edge detectors together with the map of edges true location.}
\label{fig:imtest_res_num}
\end{figure}

The proposed method is naturally robust against blur due to the above-proposed model, see Table~\ref{tbl:robustness_blurred_noise}. In fact, blurring process modifies the coefficients only in a smoother way. Of course, after having been blurred, proposed model parameters are less sensitive to edges because the intensity's discontinuities tend to spread over several pixels. However, if the blur is isotropic, pixels with the higher gradient should correspond to the edge before the blur occurs. Enlarging the neighborhood has the same impact on coefficients than if the image had been blurred, see Figure~\ref{fig:testimagec1}. The principal advantage of computing on a larger neighborhood is to deal with noise. In fact, a straightforward calculation shows that the least square model estimation of model parameters, or polynomial coefficients, have a standard deviation which is $O(\frac{\sigma}{N})$ with $\sigma$ the local noise standard deviation and $N$ the width and height of the considered neighborhood, see 
Equation~(\ref{eq:def_matrix_A_reg}). Hence, the use of a larger neighborhood offers a good way to deal with additive noise, see Table~\ref{tbl:robustness_noise}.

In order to show how the parameters $K$ and $N$ impact the robustness of the proposed edge detection method, Figures~\ref{fig:K_Influence_noise} and~\ref{fig:K_Influence_blur} show the evolution of performance as a function of Gaussian noise standard deviation and Gaussian blur radius respectively. Note that the performance ploted in those figures are the Figure of Merit which combines  missed-detection probability, false-positive rate together with location accuracy, see details in Section~\ref{sec:num_comp}.
Obviously, those two figures, calculated with the image shown in Figure~\ref{fig:imtest_FoM_Diag}, show that the robusteness increases as $K$ and $N$ increase. However, it should be noted that the test image, Figure~\ref{fig:imtest_FoM_Diag}, is very simple and thus does not allow any conclusion on the fact that increasing $K$ and $N$ also decrease the performance of the proposed edge detector for detection of edge of smaller intensity or edges near to each others.

\begin{small}
\begin{table*}[ht!]
\begin{center}
\hspace*{-3.5cm}
  \begin{tabular}{l|c|c|c|c|c}
	Detector	\!\!&\!\!	Noise std $\sigma\!=\!3$	\!\!&\!\!	Noise std $\sigma\!=\!4$	\!\!&\!\!	Noise std $\sigma\!=\!6$	\!\!&\!\!	Noise std $\sigma\!=\!8$	\!\!&\!\!	Noise std $\sigma\!=\!10$\\
					\!\!&\!\!	PSNR $\!=\!40.37$		\!\!&\!\!	PSNR $\!=\!37.88$		\!\!&\!\!	PSNR $\!=\!34.36$		\!\!&\!\!	PSNR $\!=\!31.87$		\!\!&\!\!	PSNR $\!=\!29.96$ \\
  \hline
	Canny (scale \!=\!1)  			\!\!&\!\!	$P_{FP}\!=\!8.760\%$	\!\!&\!\!	$P_{FP}\!=\!8.789\%$	\!\!&\!\!	$P_{FP}\!=\!8.989\%$	\!\!&\!\!	$P_{FP}\!=\!9.550\%$	\!\!&\!\!	$P_{FP}\!=\!10.417\%$\\
	Canny (scale \!=\!2)  			\!\!&\!\!	$P_{FP}\!=\!0.142\%$	\!\!&\!\!	$P_{FP}\!=\!0.403\%$	\!\!&\!\!	$P_{FP}\!=\!1.254\%$	\!\!&\!\!	$P_{FP}\!=\!6.109\%$	\!\!&\!\!	$P_{FP}\!=\!11.62\%$ \\
	LoG (scale \!=\!1) 		\!\!&\!\!	$P_{FP}\!=\!35.024\%$	\!\!&\!\!	$P_{FP}\!=\!36.438\%$	\!\!&\!\!	$P_{FP}\!=\!37.401\%$	\!\!&\!\!	$P_{FP}\!=\!37.784\%$	\!\!&\!\!	$P_{FP}\!=\!38.033\%$\\
	LoG (scale \!=\!2) 		\!\!&\!\!	$P_{FP}\!=\!3.758\%$	\!\!&\!\!	$P_{FP}\!=\!6.161\%$	\!\!&\!\!	$P_{FP}\!=\!14.64\%$	\!\!&\!\!	$P_{FP}\!=\!25.51\%$	\!\!&\!\!	$P_{FP}\!=\!35.18\%$ \\
	NL-filter~\cite{Laligant2010}	\!\!&\!\!	$P_{FP}\!=\!0.958\%$	\!\!&\!\!	$P_{FP}\!=\!6.139\%$	\!\!&\!\!	$P_{FP}\!=\!35.43\%$	\!\!&\!\!	$P_{FP}\!=\!\emph{\textcolor{red}{62.94\%}}$	\!\!&\!\!	$P_{FP}\!=\!\emph{\textcolor{red}{82.47\%}}$ \\
	NL-filter~\cite{Laligant2010} with median filter 	\!\!&\!\!	$P_{FP}\!=\!\emph{\textcolor{red}{39.682\%}}$	\!\!&\!\!	$P_{FP}\!=\!\emph{\textcolor{red}{43.831\%}}$	\!\!&\!\!	$P_{FP}\!=\!\emph{\textcolor{red}{48.525\%}}$	\!\!&\!\!	$P_{FP}\!=\!50.978\%$	\!\!&\!\!	$P_{FP}\!=\!52.544\%$\\
	Haralick Facet~\cite{Haralick84} \!\!&\!\!	$P_{FP}\!=\!11.364\%$	\!\!&\!\!	$P_{FP}\!=\!11.561\%$	\!\!&\!\!	$P_{FP}\!=\!11.926\%$	\!\!&\!\!	$P_{FP}\!=\!12.201\%$	\!\!&\!\!	$P_{FP}\!=\!12.483\%$\\
	Haralick Facet~\cite{Haralick84}, median filter	\!\!&\!\!	$P_{FP}\!=\!9.887\%$	\!\!&\!\!	$P_{FP}\!=\!10.128\%$	\!\!&\!\!	$P_{FP}\!=\!10.519\%$	\!\!&\!\!	$P_{FP}\!=\!10.792\%$	\!\!&\!\!	$P_{FP}\!=\!11.065\%$\\
	Proposed method	$\alpha$ ($3\times3$, order 2)\!\!&\!\!	$P_{FP}\!=\!0.143\%$	\!\!&\!\!	$P_{FP}\!=\!0.147\%$	\!\!&\!\!	$P_{FP}\!=\!0.885\%$	\!\!&\!\!	$P_{FP}\!=\!5.54\%$	\!\!&\!\!	$P_{FP}\!=\!13.52\%$ \\
	Proposed method	$\beta$ ($5\times5$, order 2)\!\!&\!\!	$P_{FP}\!=\!\mathbf{\textcolor{blue}{0.056\%}}$	\!\!&\!\!	$P_{FP}\!=\!\mathbf{\textcolor{blue}{0.057\%}}$	\!\!&\!\!	$P_{FP}\!=\!\mathbf{\textcolor{blue}{0.105\%}}$	\!\!&\!\!	$P_{FP}\!=\!\mathbf{\textcolor{blue}{0.542\%}}$	\!\!&\!\!	$P_{FP}\!=\!\mathbf{\textcolor{blue}{1.958\%}}$
  \end{tabular}
\end{center}
\caption{Numerical comparison of edge detectors robustness to white noise ; best results are shown in bold blue and worst results are shown are displayed in italic red.}
\label{tbl:robustness_noise}
\end{table*}
\end{small}

\begin{figure}[b!]
  \centering
  \def\svgwidth{0.75\textwidth}
\begingroup
  \makeatletter
    \setlength{\unitlength}{\svgwidth}
  \makeatother
  \begin{picture}(1,0.6380185)%
    \put(0,0){\includegraphics[width=\unitlength]{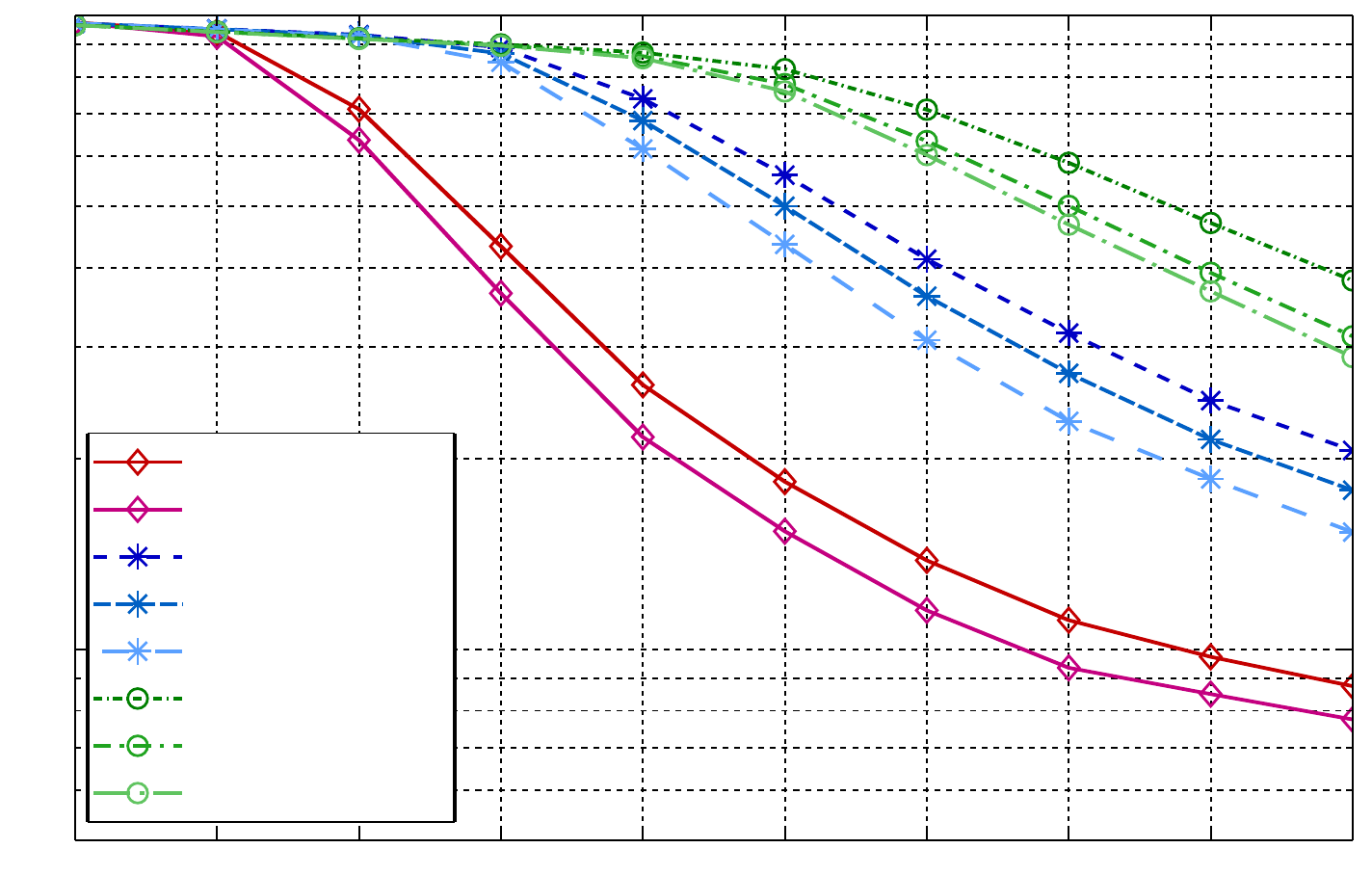}}%
\begin{footnotesize}
    \put(0.0546947,0.00050498){\color[rgb]{0,0,0}\makebox(0,0)[b]{\smash{$1$}}}%
    \put(0.15821646,0.00050498){\color[rgb]{0,0,0}\makebox(0,0)[b]{\smash{$2$}}}%
    \put(0.26173822,0.00050498){\color[rgb]{0,0,0}\makebox(0,0)[b]{\smash{$3$}}}%
    \put(0.36526212,0.00050498){\color[rgb]{0,0,0}\makebox(0,0)[b]{\smash{$4$}}}%
    \put(0.46878388,0.00050498){\color[rgb]{0,0,0}\makebox(0,0)[b]{\smash{$5$}}}%
    \put(0.57230564,0.00050498){\color[rgb]{0,0,0}\makebox(0,0)[b]{\smash{$6$}}}%
    \put(0.6758274,0.00050498){\color[rgb]{0,0,0}\makebox(0,0)[b]{\smash{$7$}}}%
    \put(0.7793513,0.00050498){\color[rgb]{0,0,0}\makebox(0,0)[b]{\smash{$8$}}}%
    \put(0.91587306,-0.00550498){\color[rgb]{0,0,0}\makebox(0,0)[b]{\smash{$\sigma$: noise std}}}%
    \put(0.04386716,0.14963224){\color[rgb]{0,0,0}\makebox(0,0)[rb]{\smash{$10^{-1}$}}}%
    \put(0.04386716,0.61175295){\color[rgb]{0,0,0}\makebox(0,0)[rb]{\smash{$1$}}}%
    \put(0.15107395,0.29224734){\color[rgb]{0,0,0}\makebox(0,0)[lb]{\smash{$N\!=\!3$ ; $K\!=\!1$}}}%
    \put(0.15107395,0.25703575){\color[rgb]{0,0,0}\makebox(0,0)[lb]{\smash{$N\!=\!3$ ; $K\!=\!2$}}}%
    \put(0.15107395,0.22358588){\color[rgb]{0,0,0}\makebox(0,0)[lb]{\smash{$N\!=\!5$ ; $K\!=\!1$}}}%
    \put(0.15107395,0.18837429){\color[rgb]{0,0,0}\makebox(0,0)[lb]{\smash{$N\!=\!5$ ; $K\!=\!2$}}}%
    \put(0.15107395,0.15316446){\color[rgb]{0,0,0}\makebox(0,0)[lb]{\smash{$N\!=\!5$ ; $K\!=\!4$}}}%
    \put(0.15107395,0.11795287){\color[rgb]{0,0,0}\makebox(0,0)[lb]{\smash{$N\!=\!7$ ; $K\!=\!2$}}}%
    \put(0.15107395,0.084503){\color[rgb]{0,0,0}\makebox(0,0)[lb]{\smash{$N\!=\!7$ ; $K\!=\!4$}}}%
    \put(0.15107395,0.04929141){\color[rgb]{0,0,0}\makebox(0,0)[lb]{\smash{$N\!=\!7$ ; $K\!=\!6$}}}%
\end{footnotesize}
  \end{picture}%
\endgroup%
\caption{Illustration of the influence of parameters $K$ and $N$: Figures of merit of the proposed test a function of Gaussian noise standard deviation $\sigma$ with blur radius set at $s=2$.}
\label{fig:K_Influence_blur}
\end{figure}

\begin{figure}[b!]
  \centering
  \def\svgwidth{0.75\textwidth}
\begingroup
  \makeatletter
    \setlength{\unitlength}{\svgwidth}
  \makeatother
  \begin{picture}(1,0.64082239)%
    \put(0,0){\includegraphics[width=\unitlength]{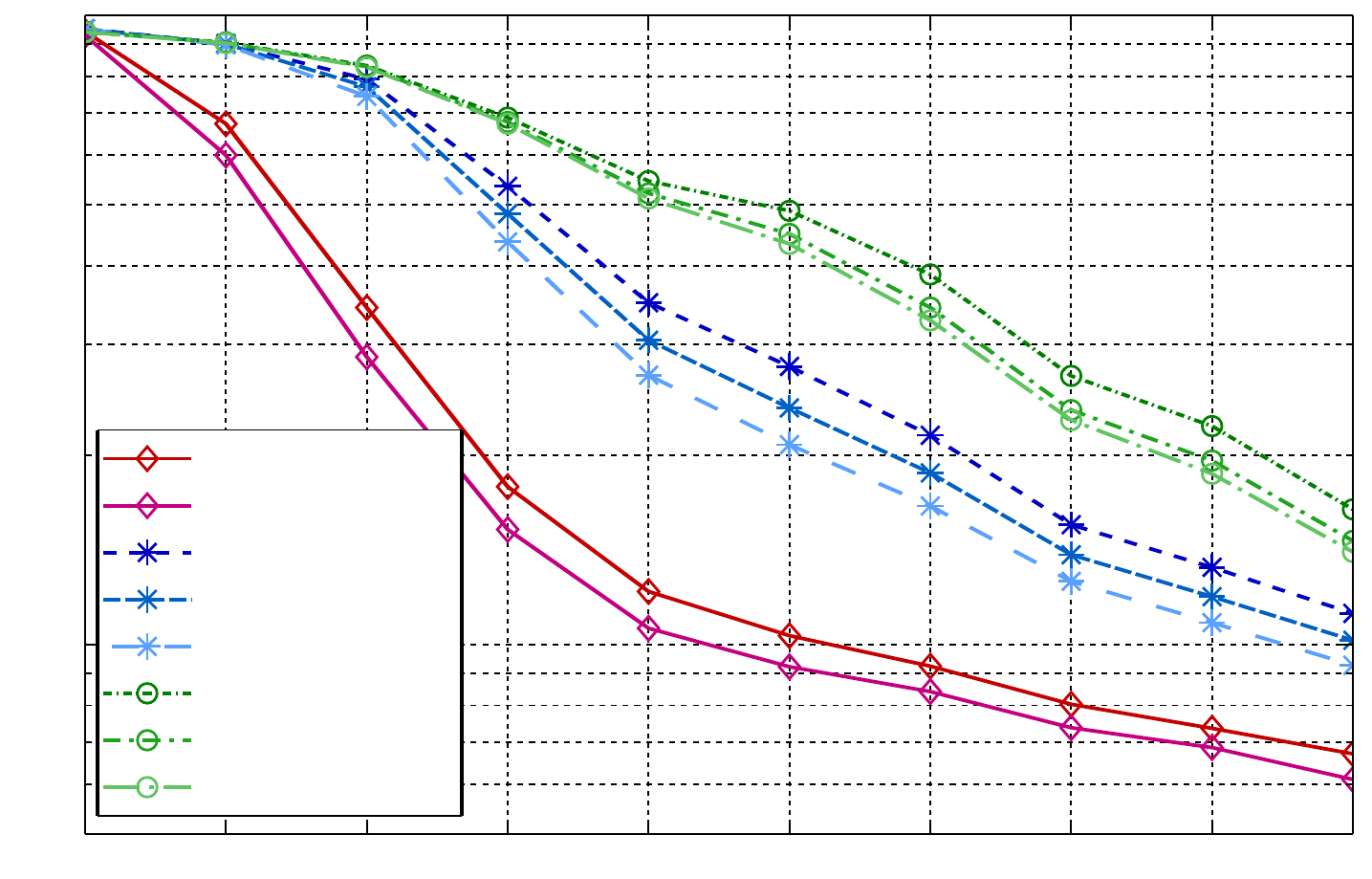}}%
\begin{footnotesize}
    \put(0.06201702,0.00050107){\color[rgb]{0,0,0}\makebox(0,0)[b]{\smash{$1$}}}%
    \put(0.1647369,0.00050107){\color[rgb]{0,0,0}\makebox(0,0)[b]{\smash{$2$}}}%
    \put(0.26745678,0.00050107){\color[rgb]{0,0,0}\makebox(0,0)[b]{\smash{$3$}}}%
    \put(0.37017879,0.00050107){\color[rgb]{0,0,0}\makebox(0,0)[b]{\smash{$4$}}}%
    \put(0.47289867,0.00050107){\color[rgb]{0,0,0}\makebox(0,0)[b]{\smash{$5$}}}%
    \put(0.57561855,0.00050107){\color[rgb]{0,0,0}\makebox(0,0)[b]{\smash{$6$}}}%
    \put(0.67833843,0.00050107){\color[rgb]{0,0,0}\makebox(0,0)[b]{\smash{$7$}}}%
    \put(0.78106044,0.00050107){\color[rgb]{0,0,0}\makebox(0,0)[b]{\smash{$8$}}}%
    \put(0.92378032,-0.00550107){\color[rgb]{0,0,0}\makebox(0,0)[b]{\smash{$s$: blur radius}}}%
    \put(0.04352736,0.15621916){\color[rgb]{0,0,0}\makebox(0,0)[rb]{\smash{$10^{-1}$}}}%
    \put(0.04352736,0.6147603){\color[rgb]{0,0,0}\makebox(0,0)[rb]{\smash{$1$}}}%
    \put(0.15764949,0.29772912){\color[rgb]{0,0,0}\makebox(0,0)[lb]{\smash{$N\!=\!3$ ; $K\!=\!1$}}}%
    \put(0.15764949,0.26279028){\color[rgb]{0,0,0}\makebox(0,0)[lb]{\smash{$N\!=\!3$ ; $K\!=\!2$}}}%
    \put(0.15764949,0.22959951){\color[rgb]{0,0,0}\makebox(0,0)[lb]{\smash{$N\!=\!5$ ; $K\!=\!1$}}}%
    \put(0.15764949,0.19466067){\color[rgb]{0,0,0}\makebox(0,0)[lb]{\smash{$N\!=\!5$ ; $K\!=\!2$}}}%
    \put(0.15764949,0.15972357){\color[rgb]{0,0,0}\makebox(0,0)[lb]{\smash{$N\!=\!5$ ; $K\!=\!4$}}}%
    \put(0.15764949,0.12478473){\color[rgb]{0,0,0}\makebox(0,0)[lb]{\smash{$N\!=\!7$ ; $K\!=\!2$}}}%
    \put(0.15764949,0.09159396){\color[rgb]{0,0,0}\makebox(0,0)[lb]{\smash{$N\!=\!7$ ; $K\!=\!4$}}}%
    \put(0.15764949,0.05665512){\color[rgb]{0,0,0}\makebox(0,0)[lb]{\smash{$N\!=\!7$ ; $K\!=\!6$}}}%
\end{footnotesize}
  \end{picture}%
\endgroup%
\caption{Illustration of the influence of parameters $K$ and $N$: Figures of merit of the proposed test a function of blur radius $s$ with standard deviation set at $\sigma=2$.}
\label{fig:K_Influence_noise}
\end{figure}
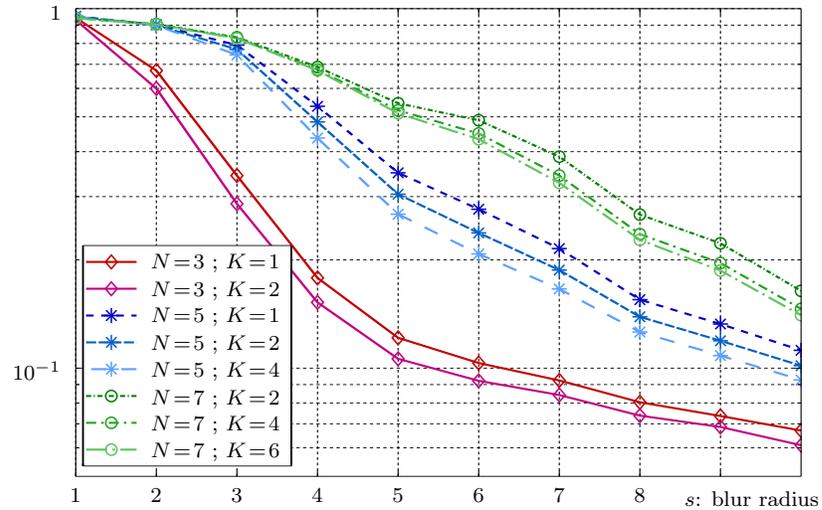

Dealing with salt and pepper noise is more problematic. This kind of noise tends to deform the image locally. Thus, this generates a high local variation $S_d$, see Equation~(\ref{eq:model_scene_decomp_disc}). The proposed model detects this noise as edges because even with blur, the local surface can not be similar to the one before noise. 
To address this problem, the use of a median filter is a usual solution which can be used together with the proposed method.

\subsection{Comparison with others methods}\label{sec:num_comp}


Measuring the accuracy and the robustness of an edge detector algorithm is a difficult  but crucial task to allow a meaningful comparison with other algorithms; this problem has been thoroughly studied in~\cite{Pratt1979,Heath96}.

In the present paper, a numerical robustness comparison of the proposed edge detection method has been made with other detectors from literature. First, it should be noted that a lot of different detectors could be used for comparison. In the present paper it is proposed to compare the proposed method with the well-known Canny detector~\cite{Canny86} and Laplacian of Gaussian (LoG) detector~\cite{Huertas86}. These two linear-filtering based detectors are known for their accuracy and their robustness. In addition, the Facet model as used for edge detection in~\cite{Haralick84} shares some similarities with the proposed method because it is based on coefficients of a local polynomial regression. Even though the method proposed in~\cite{Haralick84} exploits the polynomial coefficients to detect edge direction and to find zero crossing of directional second derivatives, which largely differs from the method proposed in this paper, the ``facet detector''~\cite{Haralick84} has been included in the numerical 
experiments due to its similarity with the proposed method. Finally, the more recent method proposed in~\cite{Laligant2010}, based on non-linear filtering, has also been used as one of the state-of-art detector.
It should also be noted that the Canny and the LoG detectors apply a (Gaussian) filter as a prior to the edge detection. Hence, for a fair comparison, the proposed edge detector as well as the Haralick's Facet method~\cite{Haralick84} and the non-linear Filter~\cite{Laligant2010} are also used with a prior filtering of inspected images. On the opposite, an hysteresis threshold procedure is used in the Canny method while a simple threshold have been applied for the other method; according to our experiments, this choice does not change the results fundamentally.

It is proposed to numerically compare edge detectors robustness in the presence of additive noise and blurring process; in fact, these two image degradations are the most commonly encountered in computer vision and thus, the most commonly considered in image processing. The methodology used in the present paper is the following. A simple test image whose edges have been previously located is used ; this test image is shown in Figure~\ref{fig:imtest_Mire} and edges are shown in Figure~\ref{fig:imtest_ground_truth}.

To measure the robustness to blurring process (followed by additive Gaussian white noise), it is proposed to calculate the detection threshold in such a manner that the total number of pixels labelled as edges corresponds to the exact number of true edge pixels. This is determined according to ground truth given in Figure~\ref{fig:imtest_ground_truth}. The probability of missed detection $P_{MD}$ is measured in percent; that is the number of pixels which belong to edge, according to the ground truth, but which are not detected by the edge detector. The results obtained using an invariant Gaussian blur with different radius are given in Table~\ref{tbl:robustness_blurred_noise}; the blurring process was followed by adding a Gaussian white noise with standard deviation $\sigma=4$.

The methodology used to measure edge detection robustness to noise is similar. It is proposed to calculate the threshold value which permits us to detect 95\% of true edges, according to ground truth provided by Figure~\ref{fig:imtest_ground_truth}. Using this threshold value the number of false-positive edge detection, denoted $P_{FP}$, that is the ---empirical--- probability to label as edge a pixel which does not belong to edge area according to ground truth. 
The results of numerical comparison are given in Table~\ref{tbl:robustness_noise}. Note that the number of false-positive is given as a percent of all the pixels which does not belong to edge area according to ground truth ; a 50\% of false positive means that half of pixels from constant areas are labeled as edge. 
Table~\ref{tbl:robustness_noise} presents the results obtained using an additive Gaussian white noise with different standard deviations, from $\sigma=3$ to $\sigma=10$. Note that for those numerical simulation a Monte-Carlo with 100 realizations has been done (for each algorithm and for each noise/blur level). 

The reason why the measure of edge detection robustness for noise and blur is slightly different is the following: when dealing with highly noised images, the main problem is to avoid the false detection of edges. On the contrary, when dealing with strong blurring process, the main difficulty is to avoid missed detection of edges.

From Tables~\ref{tbl:robustness_noise} and~\ref{tbl:robustness_blurred_noise}, it can be seen that the proposed edge detector is very robust for both noise and blur. This emphasis the relevance of the proposed methodology. For instance, results given in Table~\ref{tbl:robustness_noise} shows that the Canny edge detector ---and to a lesser extend the Haralick facet edge detector--- is very robust for noise addition while Table~\ref{tbl:robustness_blurred_noise} highlights its sensitivity to blur. On the contrary the results from Table~\ref{tbl:robustness_blurred_noise} show that the novel nonlinear edge detector proposed in~\cite{Laligant2010} have good robustness with respect to blur but is no efficient for detecting edges in the presence of white noise, see Table~\ref{tbl:robustness_noise}.

\begin{figure*}[t!]
\centerline{
  \subfloat[Test image with single diagonal edge.]{
	\includegraphics[width=0.22\textwidth, height=0.20\textwidth]{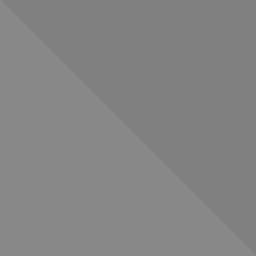}
  \label{fig:imtest_FoM_Diag}	}
\hspace{0.015\columnwidth}
  \subfloat[Ground truth with the diagonal edge of 2 pixels width.]{
	\includegraphics[width=0.22\textwidth, height=0.20\textwidth]{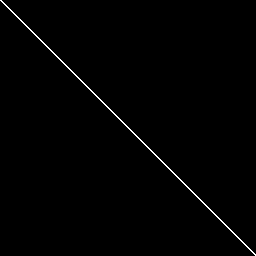}
  \label{fig:imtest_FoM_Diag_ref}	}
\hspace{0.015\columnwidth}
  \subfloat[Test ``chess'' image with horizontal and vertical edges.]{
	\includegraphics[width=0.22\textwidth, height=0.20\textwidth]{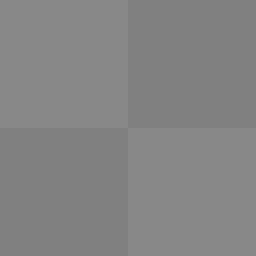}
  \label{fig:imtest_FoM_Chess}	}
\hspace{0.015\columnwidth}
  \subfloat[Ground truth with the 2 pixels width for both edges.]{
	\includegraphics[width=0.22\textwidth, height=0.20\textwidth]{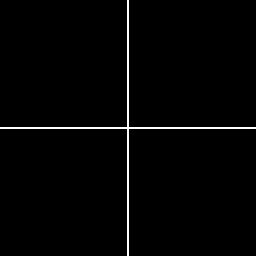}
  \label{fig:imtest_FoM_Chess_ref}	}	}
\caption{The two simple digital (before noise and blur) images used for the evaluation of edge localization and figure of merit comparison.}
\label{fig:imtest_FoM}
\end{figure*}

In addition, it is proposed to measure to accuracy of the different edge detectors in term of pixel localization. In fact, the importance of false-positive detection, that is detecting an edge where no edge is present, should be considered differently according to the distance with the nearest true edge. it is typically possible that the position of a detected edge is only one or two pixels nearby a true edge. To take into account the distance into the measure of edge detector accuracy, the so-called ``Figure of Merit'' has been proposed in~\cite{Pratt1979}, see also~\cite[Chap.5.8.2]{Jain95} and~\cite[Chap.15.5]{Pratt2007}. It is defined by the following equation~:
\begin{equation}\label{eq:FoM}
  FoM = \frac{1}{\mathrm{max}(|I_E|,|D_E|)}\sum_{(x,y) \in D_E} \frac{1}{1+\alpha d(x,y)}
\end{equation}
where $I_E$ and $D_E$ respectively represents the ``ideal'' edges (ground truth) and the detected edges, $|I_E|$ and $|D_E|$ are the cardinal of those sets (the number of pixels), $d(x,y) , (x,y)\in D_E$ is the distance between a detected edge at $(x,y)$ and the nearest true edge and $\alpha$ is a weight factor to take more or less into account the distance of falsely detected edges. The highest the figure of merit is, the more accurate the edge detector is. The figure of merit measure aims at combining probability of missed detection and false-detection together with the distance with the nearest true edge. Note that the FoM permits to compare methods between them with the \emph{Good Detection} criterion and the \emph{Good Location} criterion which are both described in \cite{Demigny2002}. Of course, the calculation of figure of merit~(\ref{eq:FoM}) is not straightforward when the ground truth exhibits many edges with complicated geometry. Hence, it is proposed in the present paper to compute the figure of merit of the compared detectors for two simple images, given in Figure~\ref{fig:imtest_FoM}. Note that the difference of intensities between the different areas is 8 and that those figures have been blurred and noised before analysis. Again a Monte-Carlo simulation with 100 repetitions, for each blurring and noise variance parameter, of figure of merit calculation have been conducted. The results obtain from this experiments are given in Figures~\ref{fig:FoM_vs_noise_std} and~\ref{fig:FoM_vs_blur}.
First, Figure~\ref{fig:FoM_vs_noise_std} shows the figure of merit of the compared detector as a function of noise standard deviation for a constant blur radius $s=2$. On the opposite, Figure~\ref{fig:FoM_vs_noise_std} shows the figure of merit of the compared detector as a function of the Gaussian blur radius for a constant noise standard deviation $\sigma=2$. Note that for a better readability only the results of edge detector used with a prior regularisation with median filtering are shown.

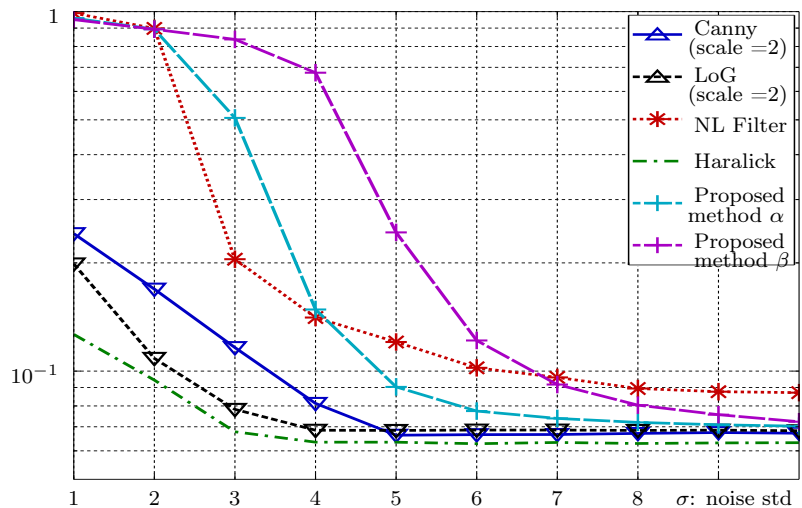
\begin{figure}[b!]
  \centering
  \def\svgwidth{0.75\textwidth}
\begingroup
  \makeatletter
    \setlength{\unitlength}{\svgwidth}
  \makeatother
  \begin{picture}(1,0.64082239)%
    \put(0,0){\includegraphics[width=\unitlength]{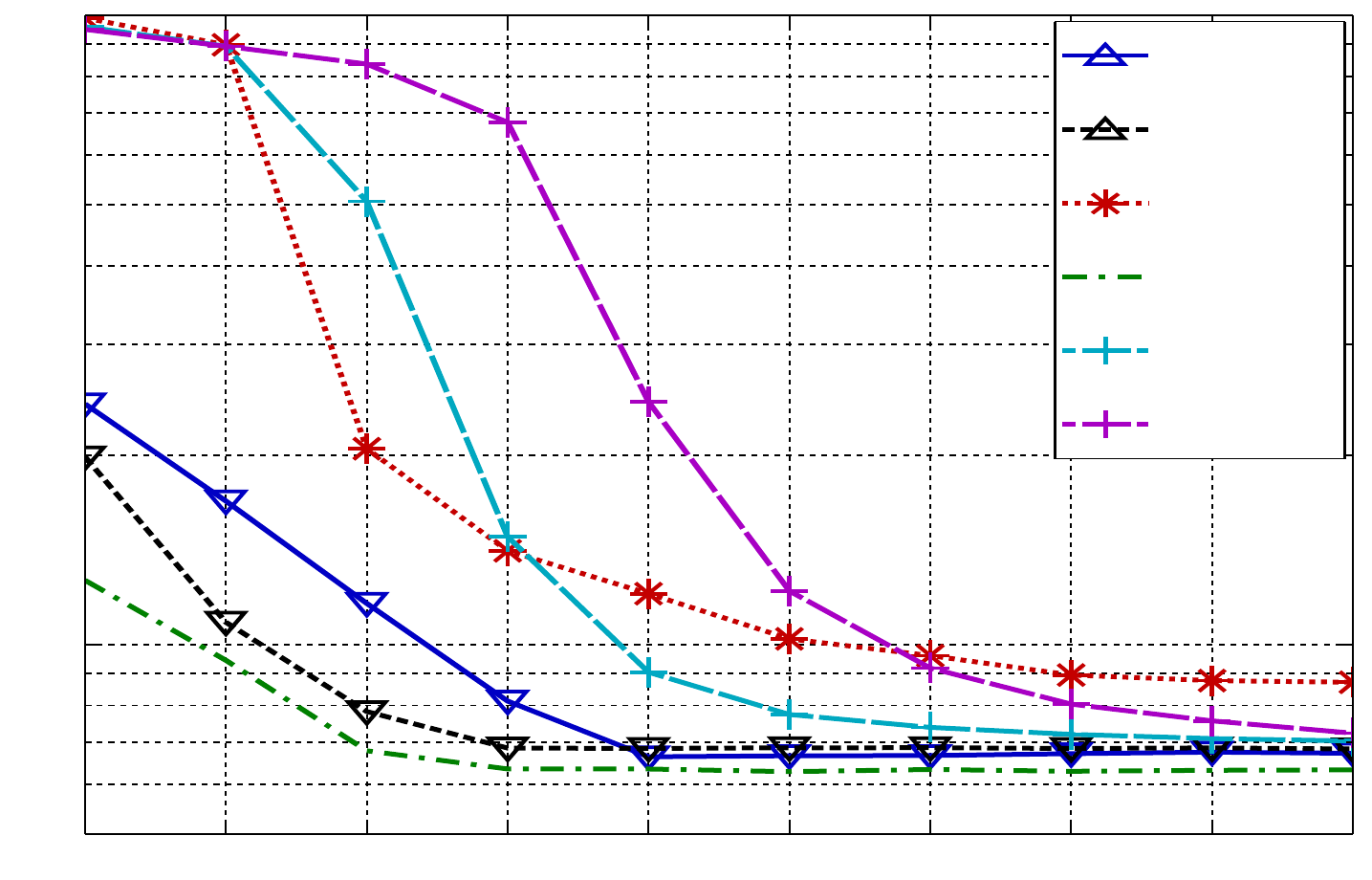}}%
\begin{footnotesize}
    \put(0.06201702,0.00050107){\color[rgb]{0,0,0}\makebox(0,0)[b]{\smash{$1$}}}%
    \put(0.1647369,0.00050107){\color[rgb]{0,0,0}\makebox(0,0)[b]{\smash{$2$}}}%
    \put(0.26745678,0.00050107){\color[rgb]{0,0,0}\makebox(0,0)[b]{\smash{$3$}}}%
    \put(0.37017879,0.00050107){\color[rgb]{0,0,0}\makebox(0,0)[b]{\smash{$4$}}}%
    \put(0.47289867,0.00050107){\color[rgb]{0,0,0}\makebox(0,0)[b]{\smash{$5$}}}%
    \put(0.57561855,0.00050107){\color[rgb]{0,0,0}\makebox(0,0)[b]{\smash{$6$}}}%
    \put(0.67833843,0.00050107){\color[rgb]{0,0,0}\makebox(0,0)[b]{\smash{$7$}}}%
    \put(0.78106044,0.00050107){\color[rgb]{0,0,0}\makebox(0,0)[b]{\smash{$8$}}}%
    \put(0.90378032,0.00050107){\color[rgb]{0,0,0}\makebox(0,0)[b]{\smash{$\sigma$: noise std}}}%
     \put(0.04352736,0.15621916){\color[rgb]{0,0,0}\makebox(0,0)[rb]{\smash{$10^{-1}$}}}%
    \put(0.04352736,0.6147603){\color[rgb]{0,0,0}\makebox(0,0)[rb]{\smash{$1$}}}%
    \put(0.85284234,0.60006269){\color[rgb]{0,0,0}\makebox(0,0)[lb]{\smash{Canny}}}%
    \put(0.84484234,0.57506269){\color[rgb]{0,0,0}\makebox(0,0)[lb]{\smash{(scale =2)}}}%
    \put(0.85284234,0.54043979){\color[rgb]{0,0,0}\makebox(0,0)[lb]{\smash{LoG}}}%
    \put(0.84484234,0.51543979){\color[rgb]{0,0,0}\makebox(0,0)[lb]{\smash{(scale =2)}}}%
    \put(0.85284234,0.47659363){\color[rgb]{0,0,0}\makebox(0,0)[lb]{\smash{NL Filter}}}%
    \put(0.85284234,0.42949792){\color[rgb]{0,0,0}\makebox(0,0)[lb]{\smash{Haralick}}}%
    \put(0.85284234,0.3868202){\color[rgb]{0,0,0}\makebox(0,0)[lb]{\smash{Proposed}}}%
    \put(0.84484234,0.3618202){\color[rgb]{0,0,0}\makebox(0,0)[lb]{\smash{method $\alpha$}}}%
    \put(0.85284234,0.32822591){\color[rgb]{0,0,0}\makebox(0,0)[lb]{\smash{Proposed}}}%
    \put(0.85284234,0.30322591){\color[rgb]{0,0,0}\makebox(0,0)[lb]{\smash{method $\beta$}}}%
\end{footnotesize}
  \end{picture}%
\endgroup
\caption{Comparison of edge detectors through figure of merit calculated for image~\ref{fig:imtest_FoM_Diag} as a function of Gaussian white noise standard deviation $\sigma$ with blur radius set at $s=2$.}
\label{fig:FoM_vs_noise_std}
\end{figure}
These two figures confirm that the Haralick facet edge detector~\cite{Haralick84} is much more sensitive to the presence of blur than white noise, see Tables~\ref{tbl:robustness_blurred_noise} and~\ref{tbl:robustness_noise}. Similarly, these figures also show that the LoG edge detector~\cite{Huertas86} is more robust than the Canny detector~\cite{Canny86} with respect to Gaussian blur but the latter performs better in the presence of noise. Finally, it can also be noted that the nonlinear-filter based edge detector~\cite{Laligant2010} remains particularly accurate is the presence of Gaussian blur, see Table~\ref{tbl:robustness_blurred_noise}, but the proposed edge detector has a better figure of merit than its competitor. This is mainly due to the fact that the false-positive edges detected by the proposed detector are, in average, nearer to true edges than the false-positive of the detector~\cite{Laligant2010}

\begin{figure}[b!]
  \centering
  \def\svgwidth{0.75\textwidth}
\begingroup
  \makeatletter
    \setlength{\unitlength}{\svgwidth}
  \makeatother
  \begin{picture}(1,0.64082239)%
    \put(0,0){\includegraphics[width=\unitlength]{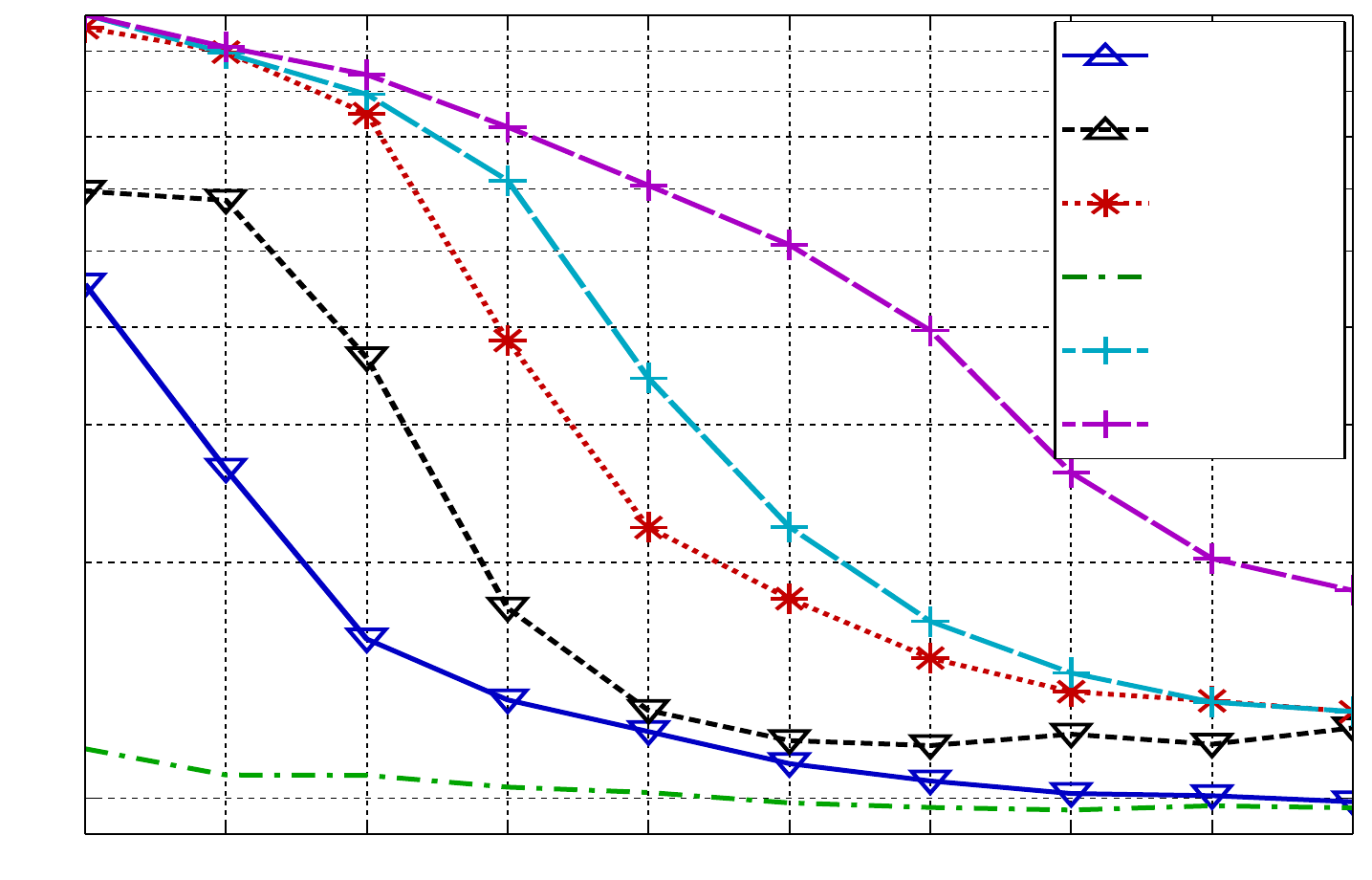}}%
\begin{footnotesize}
    \put(0.06201702,0.00050107){\color[rgb]{0,0,0}\makebox(0,0)[b]{\smash{$1$}}}%
    \put(0.1647369,0.00050107){\color[rgb]{0,0,0}\makebox(0,0)[b]{\smash{$2$}}}%
    \put(0.26745678,0.00050107){\color[rgb]{0,0,0}\makebox(0,0)[b]{\smash{$3$}}}%
    \put(0.37017879,0.00050107){\color[rgb]{0,0,0}\makebox(0,0)[b]{\smash{$4$}}}%
    \put(0.47289867,0.00050107){\color[rgb]{0,0,0}\makebox(0,0)[b]{\smash{$5$}}}%
    \put(0.57561855,0.00050107){\color[rgb]{0,0,0}\makebox(0,0)[b]{\smash{$6$}}}%
    \put(0.67833843,0.00050107){\color[rgb]{0,0,0}\makebox(0,0)[b]{\smash{$7$}}}%
    \put(0.78106044,0.00050107){\color[rgb]{0,0,0}\makebox(0,0)[b]{\smash{$8$}}}%
    \put(0.90078032,0.00050107){\color[rgb]{0,0,0}\makebox(0,0)[b]{\smash{$s$: blur radius}}}%
    \put(0.04352736,0.0442866){\color[rgb]{0,0,0}\makebox(0,0)[rb]{\smash{$10^{-1}$}}}%
    \put(0.04352736,0.6147603){\color[rgb]{0,0,0}\makebox(0,0)[rb]{\smash{$1$}}}%
    \put(0.85284234,0.60006269){\color[rgb]{0,0,0}\makebox(0,0)[lb]{\smash{Canny}}}%
    \put(0.84484234,0.57506269){\color[rgb]{0,0,0}\makebox(0,0)[lb]{\smash{(scale =2)}}}%
    \put(0.85284234,0.54043979){\color[rgb]{0,0,0}\makebox(0,0)[lb]{\smash{LoG}}}%
    \put(0.84484234,0.51543979){\color[rgb]{0,0,0}\makebox(0,0)[lb]{\smash{(scale =2)}}}%
    \put(0.85284234,0.47659363){\color[rgb]{0,0,0}\makebox(0,0)[lb]{\smash{NL Filter}}}%
    \put(0.85284234,0.42949792){\color[rgb]{0,0,0}\makebox(0,0)[lb]{\smash{Haralick}}}%
    \put(0.85284234,0.3868202){\color[rgb]{0,0,0}\makebox(0,0)[lb]{\smash{Proposed}}}%
    \put(0.84484234,0.3618202){\color[rgb]{0,0,0}\makebox(0,0)[lb]{\smash{method $\alpha$}}}%
    \put(0.85284234,0.32822591){\color[rgb]{0,0,0}\makebox(0,0)[lb]{\smash{Proposed}}}%
    \put(0.85284234,0.30322591){\color[rgb]{0,0,0}\makebox(0,0)[lb]{\smash{method $\beta$}}}%
\end{footnotesize}
  \end{picture}%
\endgroup
\caption{Comparison of edge detectors through figure of merit calculated for image~\ref{fig:imtest_FoM_Chess} as a Gaussian blur radius $s$  with white noise standard deviation fixed at $\sigma=2$.}
\label{fig:FoM_vs_blur}
\end{figure}
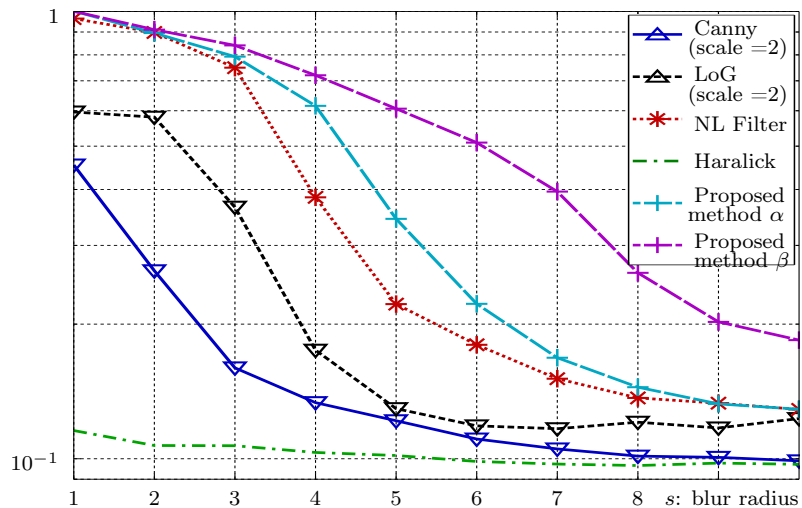

\begin{figure}[!t]
\centering
\renewcommand{\arraystretch}{0}
\hspace*{-0.3cm}
\begin{tabular}{c c}
\hspace*{-0.1cm}
\subfloat[Original Lena image.]{
\hspace*{-0.1cm}
  \includegraphics[width=0.22\textwidth]{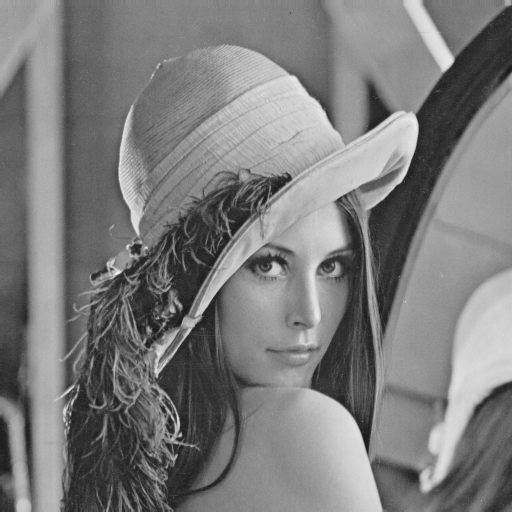}
  \label{fig:Lena_Orig} }
  \!\!\!\!&\!\!\!\!
\hspace*{0.1cm}
\subfloat[Results of Canny edge detector~\cite{Canny86}.]{
\hspace*{-0.1cm}
  \includegraphics[width=0.22\textwidth]{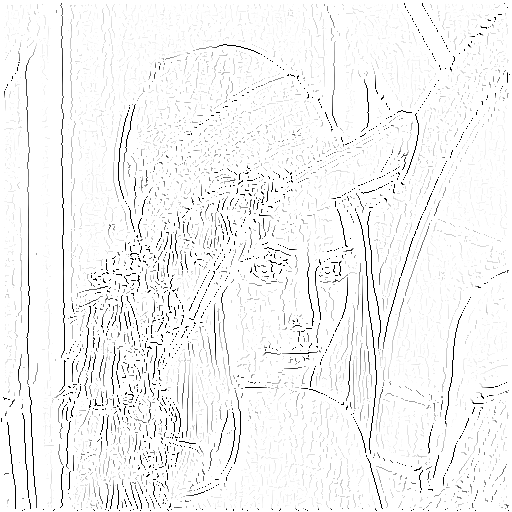}
  \label{fig:Lena_Canny} }
\\
\hspace*{-0.1cm}
\subfloat[Results of LoG edge detector~\cite{Huertas86}.]{
\hspace*{-0.1cm}
  \includegraphics[width=0.22\textwidth]{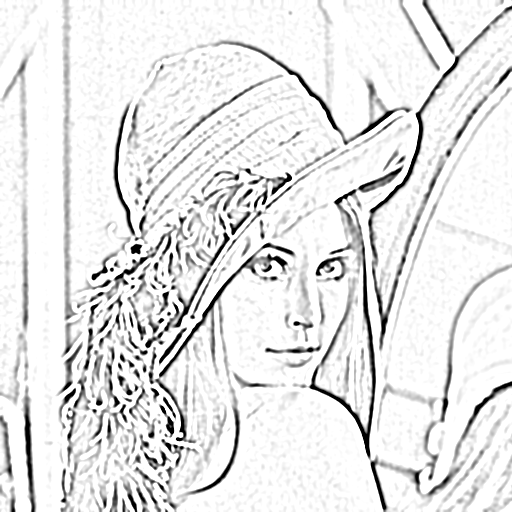}
  \label{fig:Lena_Log} }
  \!\!\!\!&\!\!\!\!
\hspace*{0.1cm}
\subfloat[Results of non-linear filtering edge detector~\cite{Laligant2010}.]{
\hspace*{-0.1cm}
  \includegraphics[width=0.22\textwidth]{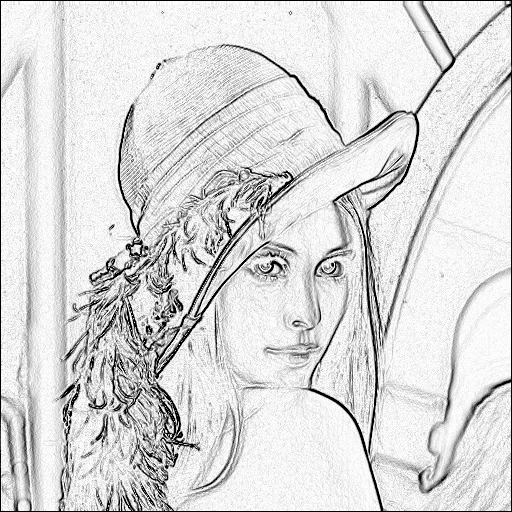}
  \label{fig:Lena_NL} }
\\
\hspace*{-0.1cm}
\subfloat[Results of Proposed edge detector with window size of $3\times3$ \& $K=2$.]{
\hspace*{-0.1cm}
  \includegraphics[width=0.22\textwidth]{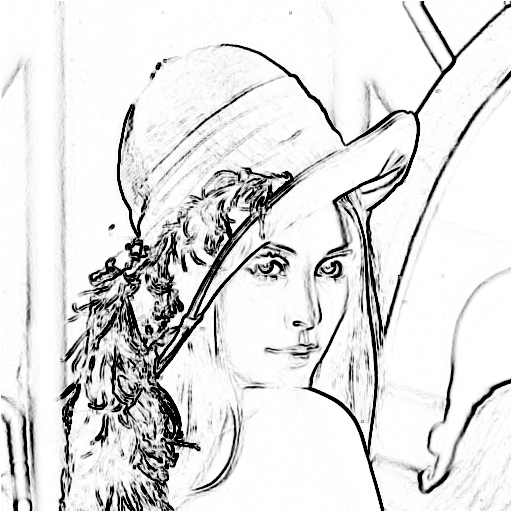}
  \label{fig:Lena_3x3} }
  \!\!\!\!&\!\!\!\!
\hspace*{0.1cm}
\subfloat[Results of Proposed edge detector with window size of $5\times5$ \& $K=2$.]{
\hspace*{-0.1cm}
  \includegraphics[width=0.22\textwidth]{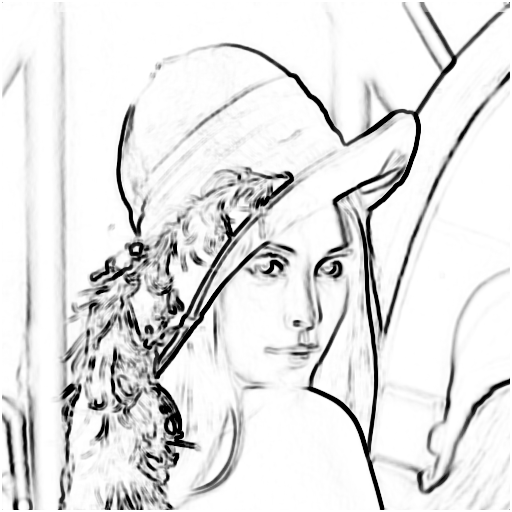}
  \label{fig:Lena_5x5} }
\end{tabular}
\caption{Comparison of edge detector result on \emph{Lena} image.}
\label{fig:Lena_EdgeDetect}
\end{figure}

%
%
%

Finally, a visual comparison of edge detectors is proposed in Figure~\ref{fig:Lena_EdgeDetect}. Those figures present the results obtained by the previously compared edge detectors with two usual images. The results obtained for the \emph{Lena} image are shown in Figure~\ref{fig:Lena_EdgeDetect}. This figure particularly shows that the edge detector proposed in~\cite{Laligant2010} is over-sensitive as it tends to highlight every single pixel which slightly differs from its neighbors. 

Similarly, the LoG~\cite{Huertas86} detector classify as edge all the pixels which differ from their neighbors, even if it is due to noise.
On the contrary, the Canny edge detector~\cite{Canny86} only coarse scale edges.

The Figure~\ref{fig:Lena_EdgeDetect} highlights this observation by using the reference image \emph{Boat} which have been subjected to Gaussian additive noise with $\sigma=10$. The results presented in Figure~\ref{fig:Lena_EdgeDetect} confirm that both the non-linear filtering edge detector~\cite{Laligant2010} and  the LoG~\cite{Huertas86} edge detector are very sensitive to local pixels with outlier value which are detected as edge. On the contrary, it should be emphasize that the proposed edge detection method provides the possibility to detect coarse to fine scale edge by modifying the parameters $N$ and $p$ which respectively represent the window size and the polynomial degree, see Figures~\ref{fig:Lena_3x3} and~\ref{fig:Lena_5x5}. In addition, by using adaptive window size, the proposed method is more robust  to additive noise than the previously proposed edge detectors as shown in~\ref{fig:Lena_3x3} and~\ref{fig:Lena_5x5}.
The results from Figures~\ref{fig:Lena_EdgeDetect} and~\ref{fig:Lena_EdgeDetect} clearly confirm the results presented in Tables~\ref{tbl:robustness_noise} and~\ref{tbl:robustness_blurred_noise} and in Figures~\ref{fig:FoM_vs_noise_std} and~\ref{fig:FoM_vs_blur}.

\section{Conclusions}
In this article, we have presented an image model which is based on a representation of a scene. The scene is considered as a juxtaposition of objects over which the physical properties of light emission vary smoothly. Taking into account the impact of the imaging system, a local model of the ensuing digital image is proposed.

Based on this model, a numerical method is proposed to discriminate discontinuities and smooth areas. The main strengths of the proposed approach are the following. First, it is not based on a prefixed pattern of searched edges of corner. Second, it offers a wide range of customizations according to the targeted goal, coarse or fine scale detection for instance. Finally, the proposed method is shown to be robust to both the blurring process and the additive noise. Some experiments have been conducted to demonstrate its efficiency. 
Our future works will aim to improve the robustness of the method here exposed and on other uses of the image model it follows from. 

\section*{Acknowledgement}

Upon acceptation, the source code used to obtained the results presented in this paper will be made avaiable.

\bibliographystyle{plain}
\bibliographystyle{model1b-num-names}
\bibliography{sigproc}

\end{document}